\documentclass[aps,jcp,twocolumn,citeautoscript,showpacs,longbibliography]{revtex4}
\usepackage{graphicx}
\usepackage{caption}
\usepackage{subcaption}
\usepackage{amsmath}
\usepackage{amssymb}
\usepackage{tabularx}
\usepackage{placeins}
\usepackage{float}	
\usepackage{units}
\usepackage{color}
\usepackage{hyperref}
\usepackage{booktabs}

\usepackage{multirow}

\usepackage{dcolumn}
\usepackage{bm}

\begin{document}

\title{New generation of effective core potentials from correlated calculations: 3d transition metal series}

\author{
Abdulgani Annaberdiyev$^{1,*}$, Guangming Wang$^{1,*}$, Cody A. Melton$^{1,2,*}$, M. Chandler Bennett$^{1,2,*}$, Luke Shulenburger$^{2}$, and Lubos Mitas$^1$}
\affiliation{
1) Department of Physics, North Carolina State University, Raleigh, North Carolina 27695-8202, USA \\
}
\affiliation{
2) Sandia National Laboratories, Albuquerque, New Mexico 87123, USA \\
}

\thanks{These authors contributed equally to this work}

\date{\today}


\begin{abstract}

Recently, we have introduced a new generation of effective core potentials (ECPs) designed for accurate correlated calculations but equally useful for a broad variety of approaches. The guiding principle has been the isospectrality of all-electron and ECP Hamiltonians for a subset of valence many-body states using correlated, nearly-exact calculations. Here we present such ECPs for the 3d transition series Sc to Zn with Ne-core, i.e, with semi-core 3s and 3p electrons in the valence space. Besides genuine many-body accuracy, the operators are simple, being represented by a few gaussians per symmetry channel with resulting potentials that are bounded everywhere. The transferability is checked on selected molecular systems over a range of geometries.  The  ECPs show a high overall accuracy with valence spectral discrepancies typically $\approx$  0.01-0.02 eV or better. They also reproduce binding curves of hydride and oxide molecules typically within 0.02-0.03 eV deviations over the full non-dissociation range of interatomic distances.

\end{abstract}

\maketitle
\section{Introduction}\label{Intro}
Effective core potentials (ECPs) and closely
related pseudopotentials provide 
a well-known technique for simplifying
electronic structure calculations
to valence-only degrees of freedom.
The ECP Hamiltonians replace the core states by potentials with projectors that mimic 
the action of the core on the 
valence electrons with different symmetries. 
For very heavy elements ECPs become almost indispensable
due to major complications that come from relativity, overwhelming energy scales of the core states and difficulties in
correlating cores in multi-atom systems.
In fact, effective core calculations might even increase 
the overall accuracy for valence properties since 
ECPs capture effects
that might be otherwise ignored. 
Clearly, one can 
adjust the ECP constructions to describe at least some of the core-core and core-valence correlation effects on the valence space as noted also 
previously \cite{Chandler:2017, DolgCao}.  
In addition, scalar relativity impacts on valence can be incorporated into ECPs in a straightforward manner. 
Furthermore, when dealing with spin-orbit explicitly ECPs conveniently conform to the transformation of
the four-component Dirac
formulation to the two-component formalism that again simplifies the 
subsequent calculations.

Over the past few decades, constructions of ECPs have evolved to a high degree of sophistication \cite{bachelet1982, tm:prb1991, vanderbilt1990, hamann2013, goedecker1996,goedecker2013,pickett1989,DolgCao}. However, most of these approaches have been focused on reproducing the original all-electron Hamiltonian within some approximate method, typically Density Functional Theory
(DFT) or Hartree-Fock(HF)/Dirac-Hartree-Fock(DHF). In
particular, the norm/shape preservation outside an effective
core radius have been used very 
extensively \cite{bachelet1982,CRENBL,SBK,LES}. 
Many constructions were also
designed to be efficient for codes with the plane wave basis. 
One of the recent focal points within the DFT framework  
has been the fidelity of 
ECPs for calculations of transition metal oxides
\cite{krogel2016}. Important 
 refinement in self-consistent 
 wave function theories 
pioneered by Stoll and coworkers 
 has been the adjustment of ECP atomic excitations  to their
all-electron HF or DHF values \cite{STU:1987,DolgCao,dolg2005}.

Further
developments have targeted better description of the electron correlations and inclusion of the many-body effects into the ECP constructions.
For example, correlated density matrices were employed as the key quantity to be matched outside the core region \cite{Acioli:1994jcp}. Another step forward has been represented 
by refinement of the ECPs that reproduce atomic excitations from correlated calculations\cite{Trail:2017jcp}. 

Recently, we have advanced the use of correlated methods in ECP constructions in a  systematic manner and we have proposed to increase the accuracy of ECP operators to a significantly higher level \cite{Chandler:2017}. This was motivated by the needs of accurate correlated methods such as Coupled Cluster (CC),
selected Configurations Interaction (CI) combined with quantum Monte Carlo (QMC) \cite{caffarel2016} and full CIQMC (FCIQMC)  \cite{booth2013} 
that allow for correlated calculations of larger systems. Indeed, the accuracy of these 
calculations started to be hampered
by the biases in the existing sets of ECPs \cite{foyevtsova2014, shulenburger:2013prb, krogel2017}
and therefore much more reliable,
accurate and tested ECPs became highly desirable. 

Let us recall 
the key principles of our correlation consistent ECP (ccECP) constructions:

i)  reproducing many-body spectrum of 
all-electron, relativistic, nearly-exact calculations for a subset of atomic states;

ii) adjusting the ECPs to capture the behavior of the original Hamiltonian 
in bonded situations for both equilibrium and non-equilibrium atomic geometries;

iii) using a simple semi-local form so that the potentials are given by a minimal or small 
set of gaussian functions;  in addition, the potentials are bounded and smooth at the origin/nucleus; we also try to keep 
the tails in nonlocal channels short so as to diminish their actions in bonding regions;

iv) and, finally, establishing a database and systematic labeling together with data on
obtained accuracy benchmarks; in addition, we keep the table open to further updates, benchmarks
and refinements.

In this work, we apply these ideas 
to the 3d transition 
metal series. These atoms present their own set 
of challenges when compared with the first- and second-rows and therefore several considerations have to be taken into account.
First, although the nominal valence levels are only 
3d and 4s, accurate ECPs require the semicore 3s and 3p states to be included
in the valence space as has been shown repeatedly in the past
\cite{STU:1987, CeperleyMitas}. This is straightforward to understand considering 
that 3s, 3p, and 3d states occupy the same electronic shell 
and therefore bonding and hybridization of 
3d levels have a significant impact
also on the semicore 
levels. Second, the elements in the middle of the 3d series are key constituents
in many magnetic materials and therefore require rather a 
precise description of related 
electronic structure properties. Often the Hund's rule high-spin atomic ground state 
 is significantly modified in the bonded environment.
This can involve several bonding mechanisms
such as charge transfer from 4s states and/or  d-s, d-p or d-d
hybridizations with resulting partial or full quench of the local magnetic moment. 
Similarly, spin flip energies and large number of states/multiplicities of the open d shell 
are important in many catalytic chemicals and materials. Particular importance for these changes in the
electronic structure is related s$\leftrightarrow$d promotion energies
as well as higher occupations of the d subshells in
both neutral and ionized cases. 

Several tables of ECPs for 3d transition metal atoms and beyond have been produced over the past 
three decades, for example, see Refs.
\cite{CRENBL, STU:1987,Burkatzki:2008jcp,Trail:2017jcp,
krogel2016}. In particular, the full periodic table developed over a period of time by the Stuttgart-Bonn-Koeln collaboration  \cite{STU:1987,dolg2005,DolgCao}
has been used quite extensively. 

We expand upon these advances here and construct high accuracy
ccECPs for the elements Sc to Zn. We use a very compact form with 
a few gaussians per channel and adjust the terms so that  
the potentials are finite
and smooth in the origin as established previously
\cite{LES,Burkatzki:2008jcp}. This form can, therefore, be used in many packages as well as with a 
variety of basis sets.
 In particular, we are interested
 in its usefulness also for periodic 
systems  such as extended 2D
materials or 3D solids calculations based on plane waves.

The all-electron and ECP calculations we use are based on scalar relativistic CC method with large basis sets and basis extrapolations. Although the calculations are
very accurate, obviously they have limitations.
The important factors are extrapolations to the 
infinite basis set limits, level of correlations in the CC method and scalar relativistic treatment. We estimate that these biases 
could sum up to $\approx$ 0.05 eV for almost all of the valence energy differences we consider. Therefore we deem 
ECPs with discrepancies within this
threshold as being of comparable accuracy. 

An additional point for future use is that we provide rather accurate values of atomic total
energies for the ground and selected excited states. The high accuracy CC method with extrapolations provide estimations
of the exact eigenvalues within systematic errors that vary from a few to about $\approx 10$ mHa for the heaviest atoms. This data offers a useful checkpoint for many methods even for those that rely on sizable error cancellations in differences such as DFT. Since practical versions of DFT employ approximate functionals this also provides a valuable data for analysis of  
the performance 
 in all-electron vs. ECP formulations
 since they can lead to non-negligible differences.

We point out
that the presented 
ECPs include also core-core and 
core-valence correlations and therefore they
represent effective Hamiltonians
that have almost as broad use as the original all-electron one, ie,  for 
many purposes they aspire to be almost {\em universal}. That should be true  essentially for any method and any valence 
property that does not require explicit core presence.
We believe that this is important because it offers a well-defined 
platform to develop a systematic understanding of biases in a variety 
of approaches. For transition
series this is particularly relevant since fully correlated relativistic all-electron calculations of such systems are out of reach for just a few atom systems.
The ECPs, therefore, enable to expand the system sizes that can be treated by many-body methods
and at the same time provide reference framework
to study systematic biases.


In what follows we first present the methods 
and objective functions used 
in the optimization process. 
The results section show the atomic and molecular properties 
of the constructed operators compared with all-electron and  other existing tables. 
We comment and discuss several aspects of these new constructions in the conclusions section. 


\section{Methods}
\subsection{ECP Parametrization}\label{ECP}
The ECP electronic Born-Oppenheimer Hamiltonian (a.u)  has 
the following form
\begin{equation}
   H_{\rm val} = \sum_i[T^{\rm kin}_i + V^{\rm pp}_i] +\sum_{i<j} 1/r_{ij}
\end{equation}
For this work, we use a semi-local ECP form with a minimal number of parameters \cite{Burkatzki:2007jcp}
\begin{equation}
    V^{\rm pp}_i = V_{loc}(r_i) + \sum_{\ell=0}^{\ell_{\rm max}} V_\ell(r_i) \sum_{m}|\ell m\rangle\langle \ell m|,
\end{equation}
where $r_i$ is the radial distance of the $i^{th}$ electron from the core's origin, and $\ell_{\rm max}$ is the maximum angular momentum for the non-local projectors. 
In this work, we choose $\ell_{\rm max} = 1$. 
The non-local terms contain the projectors on the angular momentum states with $\ell m$ quantum numbers.
The local term, $V_{loc}$, is chosen to cancel the Coulomb singularity at the origin, i.e. 
\begin{equation}
    V_{loc}(r) = -\frac{Z_{\rm eff}}{r}(1 -e^{-\alpha r^2}) + \alpha Z_{\rm eff} re^{-\beta r^2} + \sum_{i=1}^2\gamma_i e^{-\delta_i r^2},
\end{equation}
where $Z_{\rm eff}$ is the effective core charge, $Z_{\rm eff}=Z-Z_{\rm core}$.
The $V_\ell(r)$ potentials are expanded as follows 
\begin{equation}
    V_\ell(r) = \sum_{k=1}^2{\beta}_{\ell k}e^{-\alpha_{\ell k}r^2}.
\end{equation}
All variables are treated as optimization parameters in the minimization of a chosen objective function. 
In addition, a constraint that imposes a concave shape of the potential at the origin is imposed \cite{Burkatzki:2007jcp}
\begin{equation}
    \sum_{i=1}^2 \gamma_i\delta_i + \sum_{k=1}^2\alpha_{\ell k}\beta_{\ell k} > 0 , \quad \forall \ell .
\end{equation}


\subsection{Objective Function and Optimization Protocols}\label{Objective}
\begin{figure}[!htbp]
    \centering
    \caption{Spread of the contributions to the excitation energy for a variety of ECPs compared to all-electron for the Fe atom. $\Delta$HF shows the variation in the HF errors, whereas $\Delta$cVV shows the variation in the correlation energy error compared against the AE valence-valence correlation energy.}
    \label{fig:hf_vv_errors}
    \includegraphics[width=0.5\textwidth]{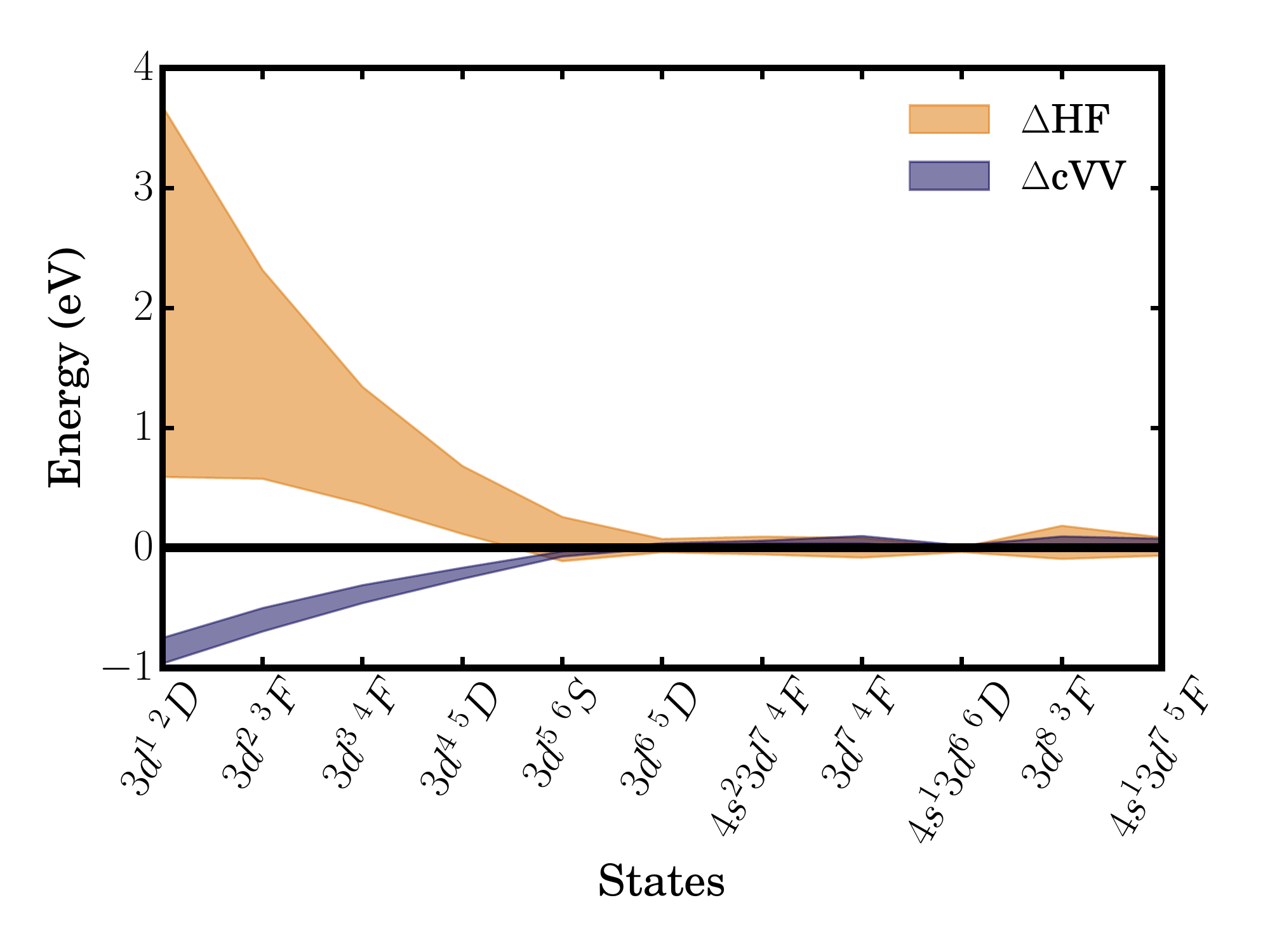}
\end{figure}
The fully correlated calculations of transition metals and heavier elements with very large basis sets and high accuracy methods such as CCSD(T) are very costly and, eventually, impractical. 
Therefore, we constructed an optimization strategy that introduces correlation into the ECP without explicitly calculating the correlation energies at each step of the optimization.
This approach relies on the fact that the correlation energy for a given atomic state is similar across different ECPs, and the variation for the correlation energy error is significantly smaller than the variation in the HF error, as shown in Figure \ref{fig:hf_vv_errors}.
For a particular excitation energy, the all-electron (AE) contribution can be written as $\Delta_g^{\rm AE} = \Delta_{\rm HF}^{\rm AE} + \Delta_{\rm CC}^{\rm AE} + \Delta_{\rm CV}^{\rm AE} + \Delta_{\rm VV}^{\rm AE}$, where CC,CV, and VV represent the core-core, core-valence, and valence-valence correlation contribution to the excitation energy.
When using an ECP, there are only valence electrons, so the ECP excitation energy can be written as $\Delta_g^{\rm ECP} = \Delta_{\rm HF}^{\rm ECP} + \Delta_{\rm corr}^{\rm ECP}$.
In order to match the AE excitation energy, the ECP can either adjust $\Delta_{\rm HF}^{\rm ECP}$ or $\Delta_{\rm corr}^{\rm ECP}$.
By considering a variety of ECPs, we plot the spread of HF errors and the difference between the ECP correlation energy and the valence-valence correlation energy of the AE in \ref{fig:hf_vv_errors} for a variety of ECPs, including BFD \cite{Burkatzki:2007jcp}, STU \cite{STU:1987}, eCEPP \cite{Trail:2017jcp}, and a variety of our constructed ECPs at both the HF and correlated levels. 
The spread of HF errors is larger than the correlation energies, which indicates that the HF contribution is more flexible across ECP parameterizations whereas the correlation energy varies much less.
This is true whether the ECP form shows smooth and finite amplitudes at the nucleus or whether it contains $-Z_{\rm eff}/r$ or $1/r^2$ divergences. 
This suggests that the $\Delta_{\rm HF}^{\rm ECP}$ can be adjusted while treating $\Delta_{\rm corr}^{\rm ECP}$ as a constant during an optimization procedure, which will result in $\Delta_g^{\rm ECP} = \Delta_g^{\rm AE}$ as is desired. 
This fact has been used in the optimization strategy below.
\begin{enumerate}
   \item For a given atom, a set of all-electron (AE) atomic states is calculated using CCSD(T) method with uncontracted aug-cc-pwCVnZ-DK \cite{balabanov2005} basis set and extrapolated to the complete basis set limit (CBS) using data for n=T,Q,5. 
For the HF, we extrapolate to the CBS limit with 
\begin{equation}
    E^{\rm HF}_n = E^{\rm HF}_{\rm CBS} + \alpha \exp\left[ -\beta n \right]
    \label{eqn:hfextrap}
\end{equation}
where $n$ labels the basis set size, $E_{\rm CBS}^{\rm HF}$ is the CBS limit, and $\alpha$ and $\beta$ are other fitting parameters.
\begin{equation}
    E^{\rm corr}_{n} = E^{\rm corr}_{CBS} + \frac{\alpha}{(n+3/8)^{3}} + \frac{\beta}{(n+3/8)^5}
    \label{eqn:corrextrap}
\end{equation}
where the correlation energy is defined as $E^{\rm corr}_n = E^{\rm CCSD(T)}_n - E^{\rm HF}_n$.
It is known that the atomic states of some transition metals such as the ground state of V are incorrectly described by a single reference using real spherical harmonics.
In order to obtain the correct atomic symmetries, we generate ``symmetry equivalenced'' orbitals by state-averaging the irreducible representations (irreps) of $D_{2h}$ for a particular atomic state \cite{symm-equiv}.
We use the resulting natural orbitals to calculate the correlation energies with CCSD(T).
This way, we are able to obtain reference AE energy gaps from a consistent description of the atomic state.


   \item The initial construction of the ECP is launched by reproducing the scalar relativistic Dirac-Hartree-Fock (DHF) spectrum.  
       We generate a large set of random ECPs and optimize each using the nonlinear DONLP2 code by Spelluci \cite{DONLP2}. 
       The objective function, $\Sigma$, to be minimized over the spectrum S is given by \\
      \begin{equation}
	  \Sigma\left(S\right) =  \sum_{s \in S} w_{s} (\Delta E_\mathrm{ECP}^{(s)}-\Delta E_{\mathrm{AE}}^{(s)})^2,
      \label{eqn:energy_consistency}
   \end{equation}
   where $\Delta E^{(s)}$ is the excitation energy for state $s$ relative to the neutral ground state and $w_{s}$ is the weight assigned to that excitation.
   The spectrum is optimized using numerical DHF code \cite{shirley} to avoid basis set errors and to speed up the calculations.
   The set of states considered includes the neutral and singly ionized s$ \leftrightarrow$d excitations, as well as further ionizations and the anion ground state. 
   These states are chosen due to the fact that the low-lying s$\leftrightarrow$d transitions and various oxidation states play a crucial role in transition metal chemistry \cite{STU:1987}. 
   The resulting ECPs are further refined as follows. 

   \item 
   The DHF spectrum optimized ECP with the least objective function value is used to calculate the correlation energies for the states mentioned above using the same method described for the AE case.
   \item It is clear that DHF spectrum optimized ECP atomic excitation energies $\Delta E_\mathrm{ECP}^{(s)}$ are not going to match the corresponding AE excitation energies $\Delta E_{\mathrm{AE}}^{(s)}$ when a post-HF method such as CCSD(T) or CI is used. However, it is possible to obtain $\Delta_g^{\rm ECP} = \Delta_g^{\rm AE}$ by only matching $\Delta_{\rm HF}^{\rm ECP}$ to \emph{shifted} $\Delta_{\rm HF}^{\rm AE}$:
   \begin{equation}
   \Delta_{\rm HF}^{\rm ECP} = \Delta_{\rm HF}^{\rm AE} + \epsilon
   \end{equation}

    where $\epsilon$ is given as:\\
       \begin{equation}
       \begin{aligned}
       \epsilon & = \Delta^{\rm AE}_{\rm corr}-\Delta^{\rm ECP}_{\rm corr} \\
       & = (E^{\rm AE}_{\rm corr,excited}-E^{\rm AE}_{\rm corr,ground})- \\
       & \quad (E^{\rm ECP}_{\rm corr,excited}-E^{\rm ECP}_{\rm corr,ground})
       \end{aligned}
       \end{equation}
    Here AE correlation energy refers to the total correlation energy (CC+CV+VV) of the state.
    The new shifted gaps are inserted into the objective function in equation \ref{eqn:energy_consistency} to be minimized. 
    
    \item The steps 3-4 are iterated until a self-consistent ECP is obtained. For each step $\epsilon$ is re-evaluated using the ECP from the last iteration. 
    In every iteration, the ECP parameters are randomly perturbed within $\approx$ 1-2\% of parameter values to ensure better scanning of possible local minimas around the current values of the parameters. 
    Usually, a set of self-consistent parameters are obtained within a few iterations.

    In cases where pure energy consistency does not result in an acceptably transferable ECP (as described in section \ref{Results}), we reduce the spectrum by removing very highly ionized states and add additional constraints to improve the transferability. Note that the full spectral fits include also core-core and core-valence
correlations that become quite significant 
for the highly ionized states and therefore
they have tendency to steer the ECP 
from the optimal 
valence-valence description. This is especially true for our ECP form with 
small number of variational parameters. Although the bias is typically small 
(say, 0.1 eV) we 
opted for further refinements. This is 
accomplished by including one-particle 
eigenvalue discrepancies into the objective function. It directs 
the highly ionized states to be less 
influenced by the corresponding 
correlations so as to be increase
the weight of the HF character for these states.
In these cases, we utilize a new objective function, $\Gamma$, given as 
\begin{equation}
    \Gamma\left(S\right) = \Sigma(S) + \gamma \sum_{\ell}\limits\left( \epsilon_\ell^{\rm ECP}-\epsilon_{\ell}^{\rm AE}\right)^2
    \label{eqn:final_objective}
\end{equation}
where the first term is the initial spectral objective function and $\epsilon_{\ell}$ is one particle eigenvalue.
The eigenvalues are weighted by $\gamma$ to allow the spectrum to be minimized while keeping the one-particle eigenvalues reasonably close to the corresponding AE ones. With few iterations one can find a 
compromise that reproduces a large
part of the spectrum as well achieves transferability in molecular calculations.
Clearly, further improvements are possible but in this work we are mostly concerned with the demonstration of principle and with providing simple ECPs that fulfill the accuracy criteria.

\end{enumerate}

\section{Results}\label{Results}
\begin{figure*}[!htbp]
    \begin{subfigure}{0.5\textwidth}
        \centering
        \includegraphics[width=\textwidth]{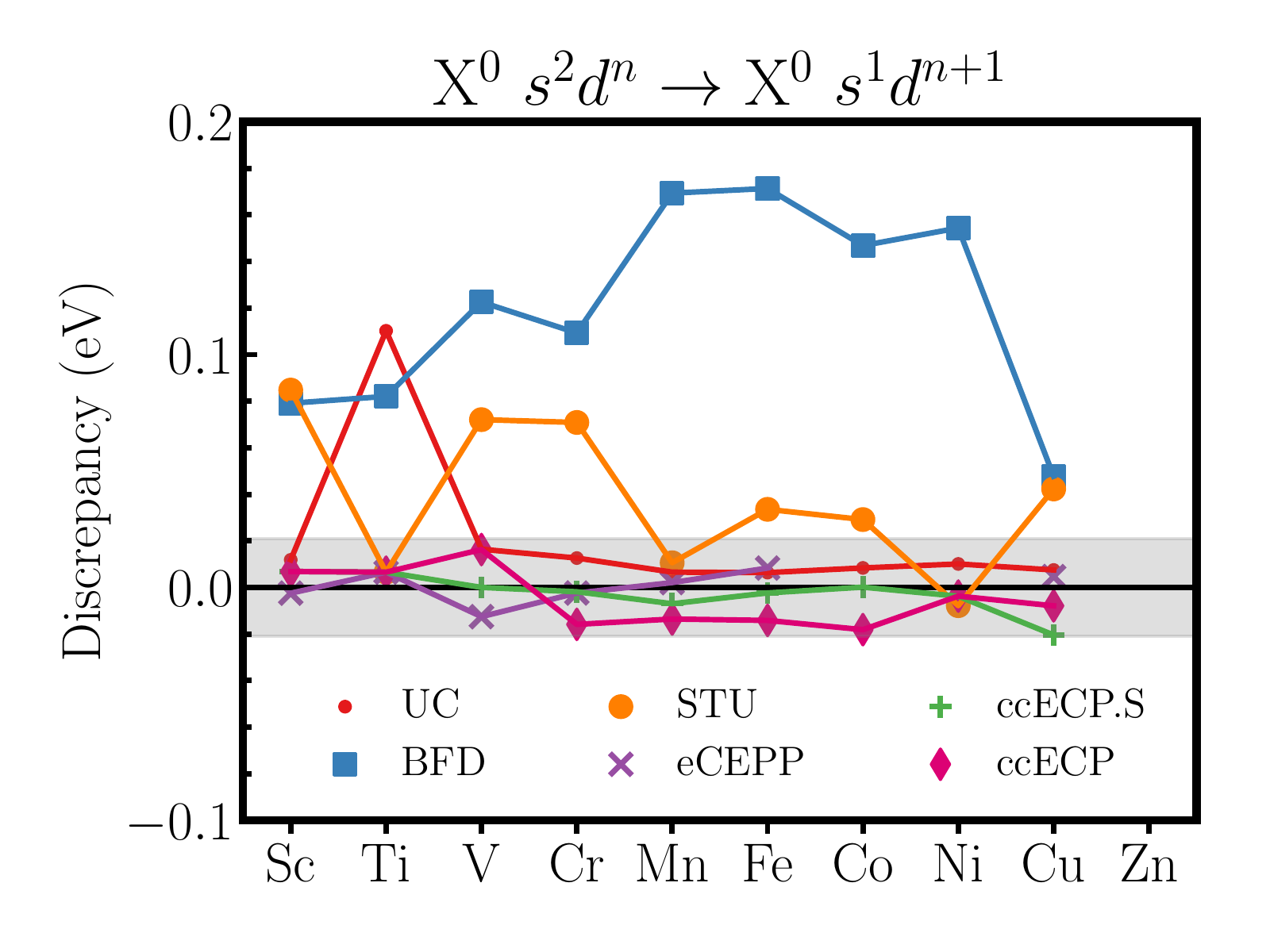}
	\caption{}
        \label{fig:Is2dn_to_Is1dn1}
    \end{subfigure}%
    \begin{subfigure}{0.5\textwidth}
        \centering
        \includegraphics[width=\textwidth]{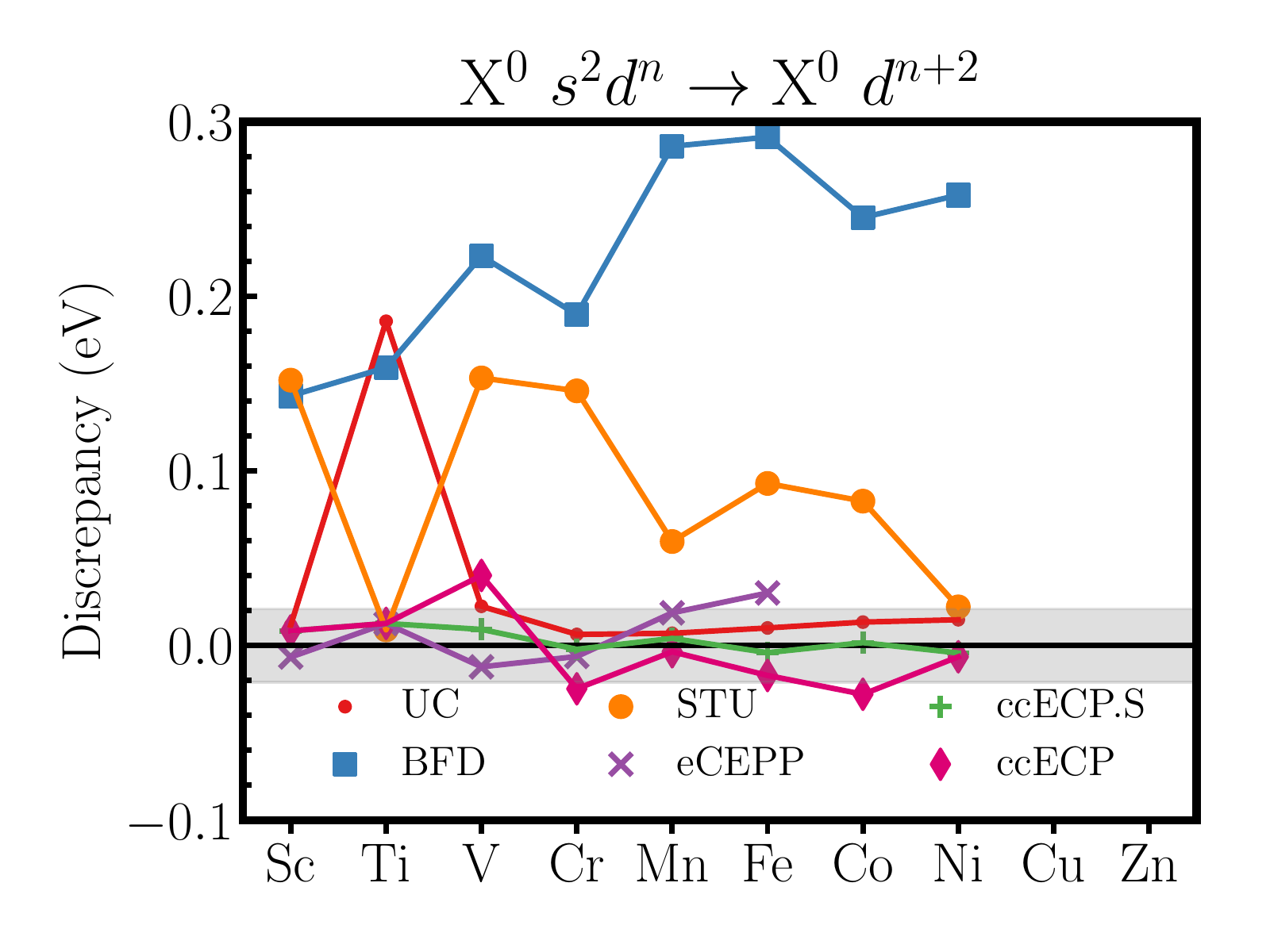}
	\caption{}
        \label{fig:Is2dn_to_Idn2}
    \end{subfigure}
    \begin{subfigure}{0.5\textwidth}
        \centering
        \includegraphics[width=\textwidth]{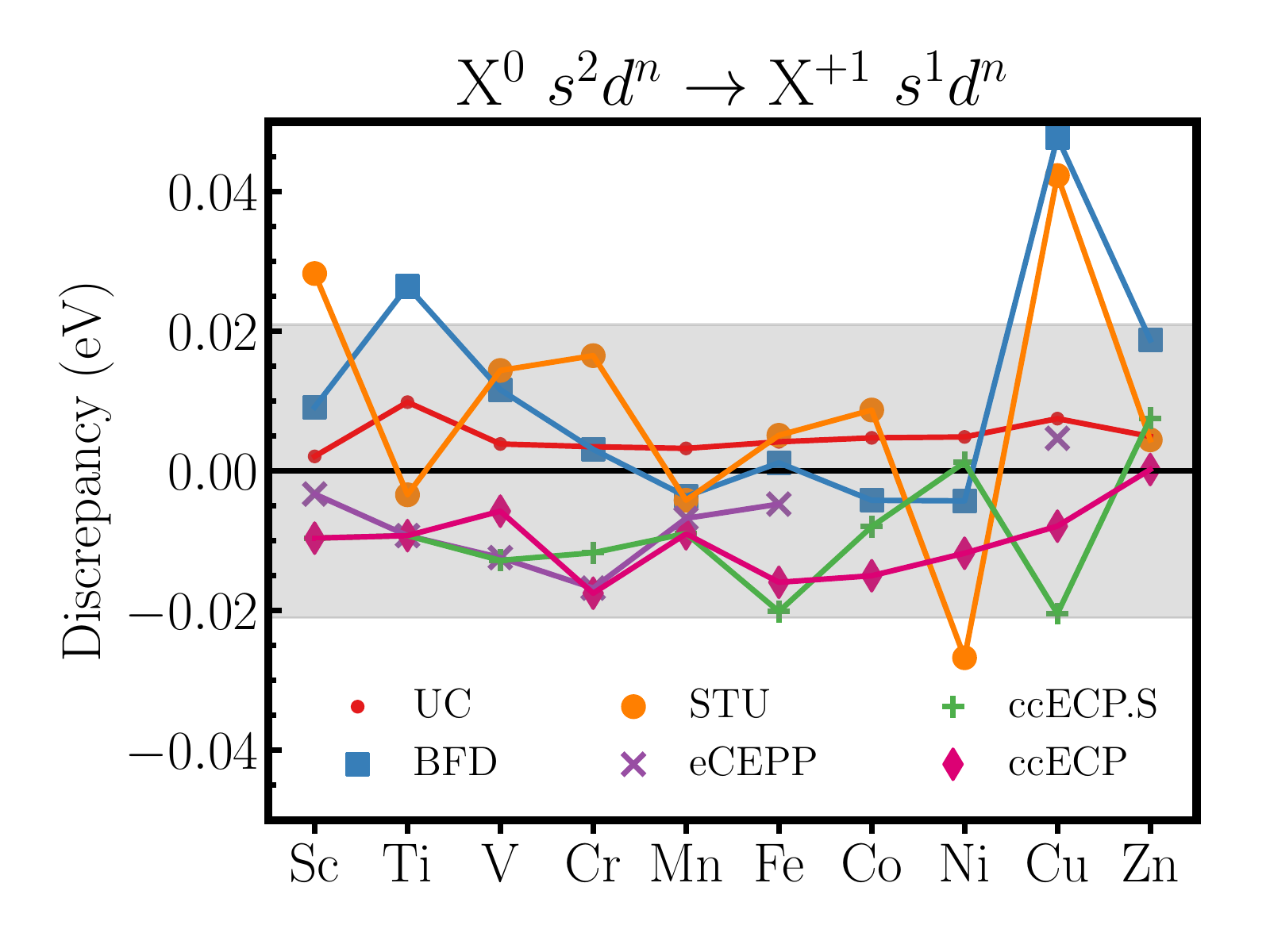}
	\caption{}
        \label{fig:Is2dn_to_IIs1dn}
    \end{subfigure}%
    \begin{subfigure}{0.5\textwidth}
        \centering
        \includegraphics[width=\textwidth]{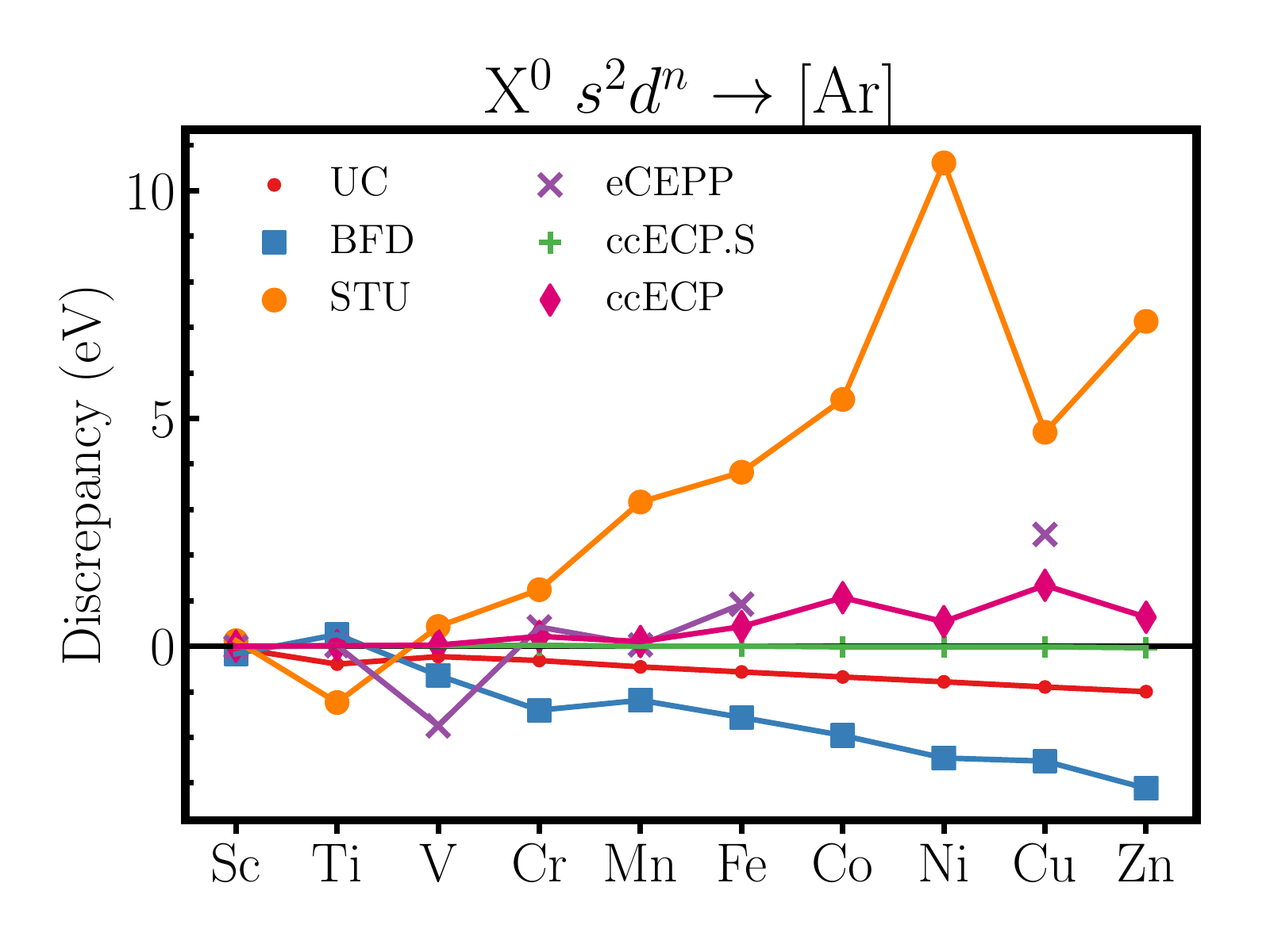}
	\caption{}
        \label{fig:Is2dn_to_Ar}
    \end{subfigure}
    \caption{Discrepancies for the various core approximations compared to all-electron CCSD(T) for select states. Each state discrepancy uses the neutral $s^2d^n$ occupation as the reference, which is the neutral ground state for each transition metal except for Cr and Cu, which have $s^1d^5$ and $s^1d^{10}$ ground states correspondingly. (a) the neutral $s^2d^n\rightarrow s^1d^{n+1}$ excitation. (b) the neutral $s^2d^n \rightarrow d^{n+2}$ (c) the ionization from $s^2d^n \rightarrow s^1d^n$ (d) the ionization from $s^2d^n \rightarrow [Ar]$. The shaded gray window in each figure indicates a discrepancy of half of chemical accuracy in either direction from the all-electron reference. We note that for Sc and Ti, our final ccECP is equivalent to our spectral ccECP.S.}
    \label{fig:states}
\end{figure*}

\begin{figure}[!htbp]
    \centering
    \caption{Mean Absolute Deviation, or MAD, for the TMs considering a large part of the spectrum from [Ar] up to low-lying neutral excited states and the anion. The shaded region of half of chemical accuracy is not visible on this scale. We note that for Sc and Ti, our final ccECP is equivalent to our spectral ccECP.S}
    \label{fig:mad}
    \includegraphics[width=0.5\textwidth]{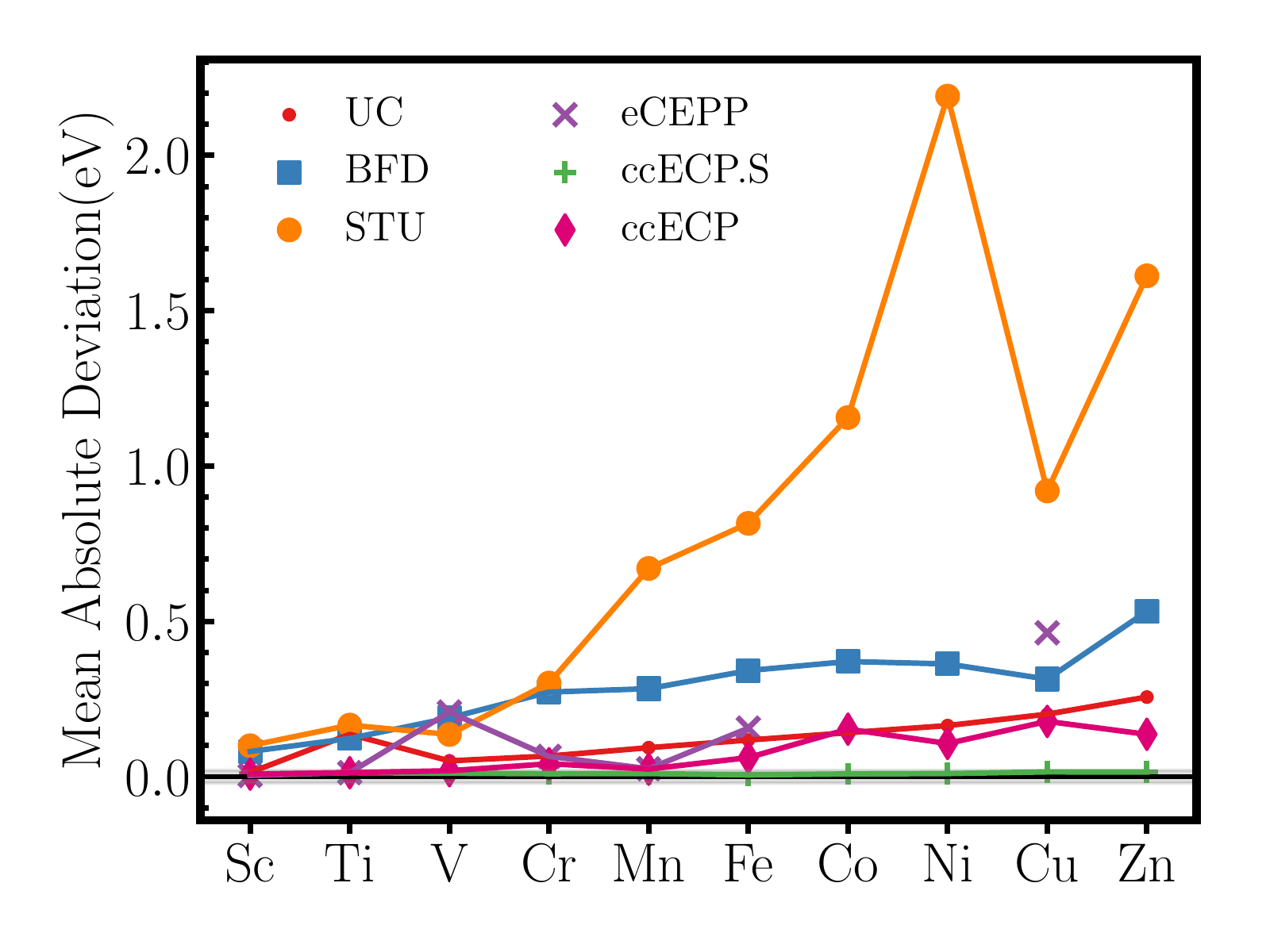}
\end{figure}

We present results comparing our correlation consistent ECPs, labeled ccECP, to various other core approximations relative to the all-electron calculation for both the atomic spectra as well as monohydrides and monoxides. 
The all-electron reference in all following calculations is spin-restricted CCSD(T) where we have correlated all electrons in the system, i.e. no core orbitals have been frozen. 
To include relativity, we utilize a 10$^{th}$ order Douglass-Kroll-Hess Hamiltonian \cite{Reiher:2004jcp}. 
To fully correlate the core, we use an uncontracted aug-cc-pwCV$n$Z-dk basis \cite{balabanov2005} and extrapolate to the complete basis set limit to minimize basis set error.
We compare our ECPs against an all-electron {\rm uncorrelated} core, which does not allow for any excitations from the Neon core in the CCSD(T), as well as other ECPs.  
The uncorrelated core approximation, labeled UC, only accounts for valence-valence correlation, and any core-core or core-valence correlation is absent despite it being an all-electron calculation.
The other ECPs include Burkatzki-Filippi-Dolg ECPs (BFD) \cite{Burkatzki:2007jcp} which are energy-consistent DHF ECPs designed for use in QMC, Stuttgart ECPs (STU) \cite{STU:1987} which are energy-consistent DHF ECPs, and the recently constructed Trail-Needs ECPs (eCEPP) \cite{Trail:2017jcp} which use a shape and energy consistent scheme to construct correlated ECPs.

Obviously, the methods we use have their 
accuracy limits.
In particular, we rely on
scalar relativistic approach with averaged 
spin-orbit effects. This introduces additional
bias of $\approx$ 0.025 eV as 
a representative value that we  estimated 
from accurate spin-orbit correlated calculations of atomic excitations\cite{spin_orbit}.
Further important sources of bias are the sizes of basis sets (alleviated by extrapolations)
and level of correlation captured by the CCSD(T) method. On smaller
systems where we were able to push the basis set limits up to $n=6,7$, ie, 6Z,7Z and level of correlation up to CCSDT(Q) level (triples 
explicitly and quadruples perturbationally). These limited calculations suggest that it is difficult to ascretain accuracies better than rougly 0.02 eV
for energy differences. In addition, for the heaviest atoms, we were not able to carry out the full sequence of basis set calculations up to 5Z, so this threshold is probably mildly higher. Clearly, 
ECPs with discrepancies below these inherent biases of 
$\approx$ 0.05 eV
(that include also a number of our
constructions presented below) 
are certainly interesting for methodology reasons. However, 
 the ECPs within such bound should be considered as being of comparable quality.

We briefly discuss the results for every element in the following subsections.
In section \ref{Atomic Spectra}, we present an overview of the atomic spectra for all transition metals. 
In sections \ref{sec:Sc}-\ref{sec:Zn}, for every atom we show the atomic spectrum discrepancies and transferability tests on monohydrides and monoxides. In these molecules, we test various core approximations near and out of equilibrium geometries, i.e., compressed bond lengths. 
In general, we show results for our spectral-only ECP (labeled ccECP.S) as well as the ECP that compromises a small part of the spectrum for increase in the overall transferability (this is labeled as ccECP).
For oxygen ECP, we use the ECP from the corresponding table; our oxygen ccECP can be found in our previous work\cite{Chandler:2017}.
Although for some of the oxide molecules  we were unable to converge with 5Z basis sets
due to technical difficulties. 
However, we note that in terms of the binding curves discrepancies  are well converged at the VQZ level.
Therefore, for each molecule we show only the largest basis set that we converged as opposed to the CBS extrapolation.

All ECP parameters are given in Table \ref{tab:1st_param} and Table \ref{tab:2nd_param} while our hydrogen ECP is given in Table \ref{tab:pp_H}.


\subsection{Atomic Spectra}\label{Atomic Spectra}

We first provide an overview of the atomic spectra.
The optimized spectrum for ccECP.S included the neutral $s^2d^n$, $s^1d^{n+1}$, $d^{n+2}$, the first ionized $s^1d^n$ and $d^{n+1}$, all ionizations down to the Ar core, and the [Ne]$3s^2$ state.
As we will see in future sections, the inclusion of these highly ionized states can lead to non-negligible biases and a decrease in the overall transferability. 
Therefore, in cases where this full spectrum is insufficient, we add in constraints and work with a reduced spectrum as described in section \ref{Objective}.

Figure \ref{fig:states} shows the errors across the entire transition metal series for select atomic states ranging from some low lying excited states to a highly ionized state. 
In Figs. \ref{fig:Is2dn_to_Is1dn1} and \ref{fig:Is2dn_to_Idn2}, we show the neutral s$\leftrightarrow$d transitions from the $s^2d^n$ to $s^1d^{n+1}$ and $d^{n+2}$, respectively.
Both STU and BFD, result in discrepancies on the order of 0.1 eV throughout the entire transition metal series for these low lying atomic states. 
UC, eCEPP, and our ccECP.S and ccECP all result in discrepancies within half of the chemical accuracy, indicated by the shaded region. 
Figure \ref{fig:Is2dn_to_IIs1dn} shows the $s^2d^n$ to $s^1d^n$ ionization, which is well described with all core approximations. 
It is important to note that STU and BFD include many of these states directly into their optimization as well. 
However, the neglect of directly introducing correlation into the construction can result in large errors in the atomic spectra, despite being well optimized at the HF/DHF level as illustrated by both the eCEPP and our ccECPs. 

As the ionization level is increased, all previous core approximations begin to show larger discrepancies with the AE reference. In order to 
illustrate this, we show the ionization down to the Ar core in Figure \ref{fig:Is2dn_to_Ar}. 
The STU and BFD ECPs we see significantly larger discrepancies, ranging from 1-10~eV throughout the series.  
The correlated eCEPPs show an improvement but still result in discrepancies of a few eV. 
The all-electron UC approximation results in a slightly increasing error as the atomic number increases up to roughly 1~eV for Zn. 
Our energy consistent ECPs (ccECP.S), however, are able to maintain a uniform accuracy within half of the chemical accuracy for each atomic state and for all ionizations. 

In Figure \ref{fig:mad}, we show the mean absolute deviation,
\begin{equation}
    {\rm MAD} = \frac{1}{N_s} \sum\limits_{s=1}^{N_s} \left| \Delta_s^{\rm ECP} - \Delta_s^{\rm AE} \right|
\end{equation}
across all of the states including the low-lying s$\leftrightarrow$d transitions, anions, and ionizations down to the semicore [Ar]. 
Our ccECP.S ECPs are able to maintain MADs of nearly 0.01~eV for all transition atoms. 
In the following sections focused on the individual atoms, we investigate energy consistency 
influences the transferability of the ECPs for each element. 
Our adjusted ccECP constructions, which slightly compromise the atomic spectrum when compared to the ccECP.S (as illustrated in Fig. \ref{fig:mad}), still maintain higher accuracy for the MADs when compared against the other tested ECP constructions. 

\subsection{Scandium}\label{sec:Sc}

\begin{table*}[!htbp]
\centering
\caption{Sc AE gaps and relative errors for various core approximations. All gaps are relative to the [Ar] $3d^1 4s^2$ $^2D$ state. LMAD represents the mean absolute deviation of low-lying gaps (EA, IP and $s \leftrightarrow d$ ). MAD is mean absolute deviation for all gaps. All values in eV.}
\label{tab:Sc}
\begin{tabular}{ll|rrrrrrrrr}
\hline\hline

Gaps &  &          AE &        UC &       BFD &       STU &     eCEPP &   ccECP$^\dagger$\\
\hline
{[Ar] $3d^24s^2$ } &      $^3F$ &     0.509479 &     0.003048 &   0.043130   &     0.064818 &    -0.000680 &    0.008299  \\
{[Ar] $3d^24s^1$ } &      $^4F$ &     1.419374 &     0.011864 &   0.079103   &     0.084764 &    -0.002449 &    0.006857  \\
{[Ar] $3d^3$ }     &      $^4F$ &     4.283618 &     0.011646 &   0.143078   &     0.152030 &    -0.006531 &    0.008463  \\
{[Ar] $3d^14s^1$ } &      $^3D$ &     6.542273 &     0.002095 &   0.009143   &     0.028273 &    -0.003293 &   -0.009606  \\
{[Ar] $3d^2$ }     &      $^3F$ &     7.144951 &     0.019375 &   0.112982   &     0.138533 &    -0.000463 &   -0.000408  \\
{[Ar] $3d^1$ }     &      $^2D$ &    19.344148 &     0.004626 &   0.015048   &     0.105471 &     0.001551 &   -0.021660  \\
{[Ar]}             &      $^1S$ &    44.069078 &    -0.040491 &  -0.166016   &     0.123866 &    -0.013715 &    0.002667  \\
\hline
LMAD & & & 0.009606& 0.077487& 0.093684& 0.002683& 0.006727\\
MAD & & &  0.013306& 0.081214& 0.099679& 0.004097& 0.008280\\
\hline\hline
\end{tabular}
\\{\small $^\dagger$ We note that our ccECP is equivalent to our spectral construction, ccECP.S as described in the text}
\end{table*}

\begin{figure*}[!htbp]
\centering
\begin{subfigure}{0.5\textwidth}
\includegraphics[width=\textwidth]{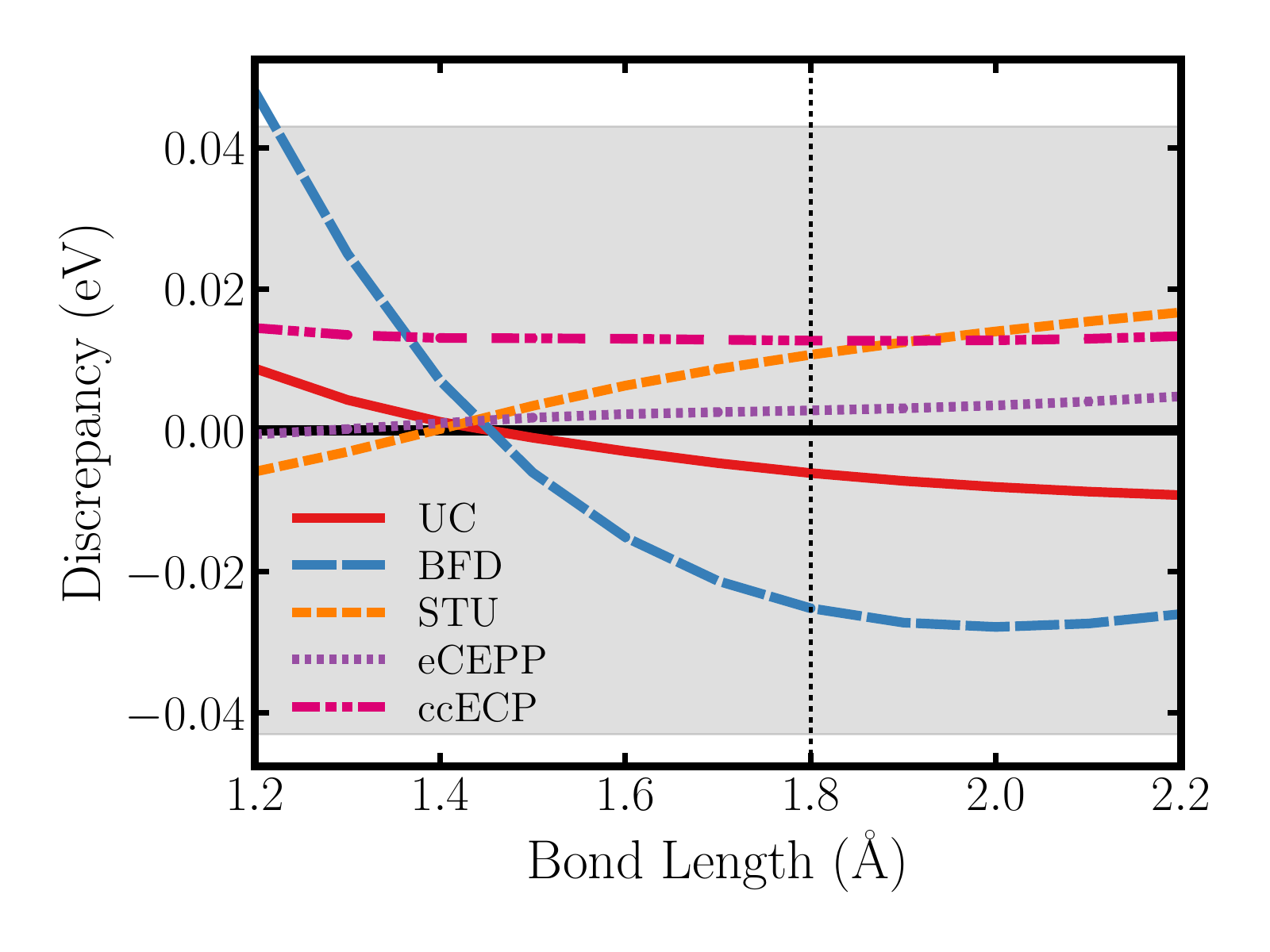}
\caption{ScH binding curve discrepancies}
\label{fig:ScH}
\end{subfigure}%
\begin{subfigure}{0.5\textwidth}
\includegraphics[width=\textwidth]{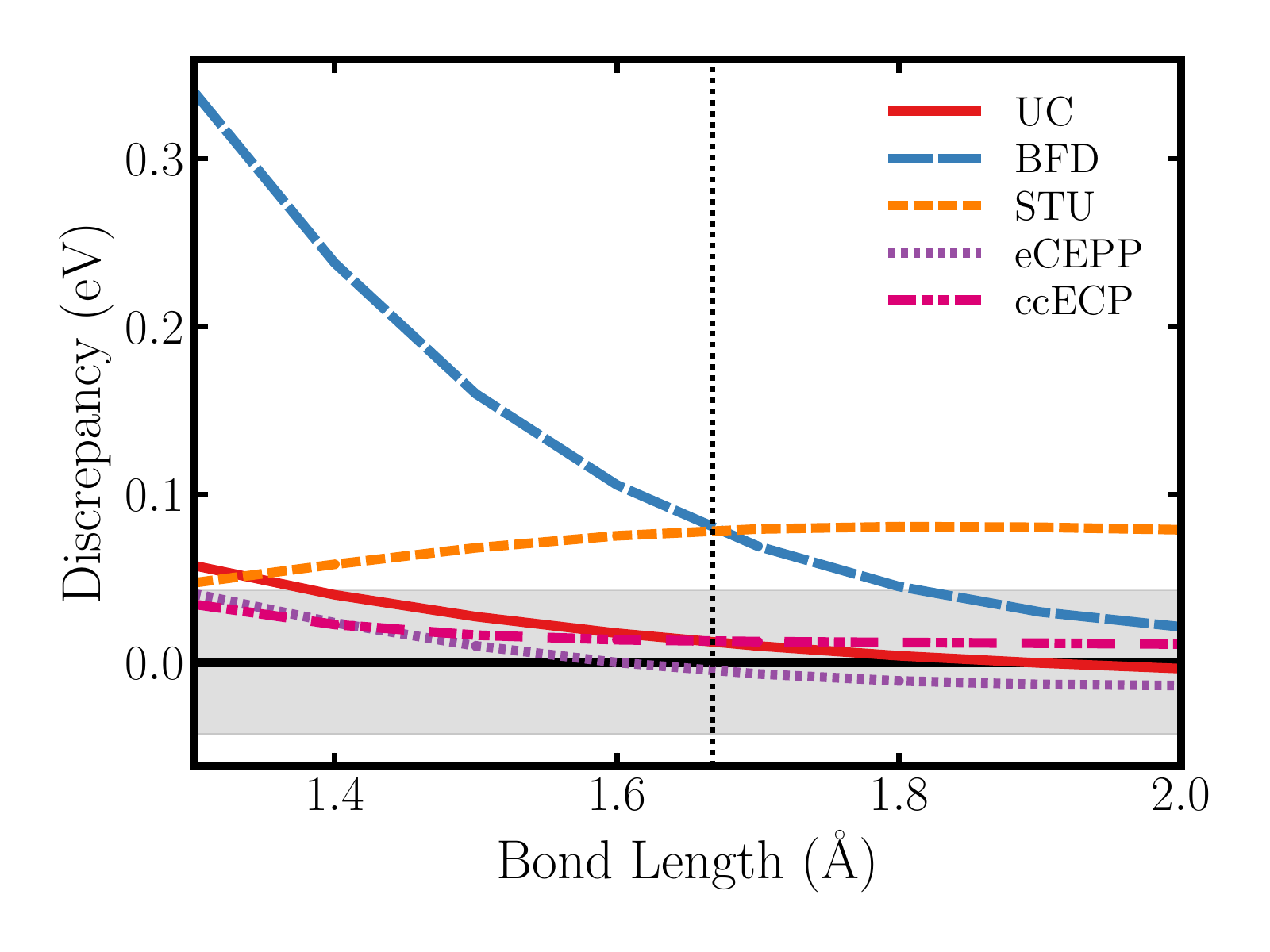}
\caption{ScO binding curve discrepancies}
\label{fig:ScO}
\end{subfigure}
\caption{Binding energy discrepancies for (a) ScH and (b) ScO molecules. The binding curves are relative to the AE CCSD(T) binding curve. The shaded region indicates a discrepancy of chemical accuracy in either direction. }
\label{fig:Sc_mols}
\end{figure*}

In the case of Sc, the atomic and molecular data is given in Table \ref{tab:Sc} and Fig. \ref{fig:Sc_mols}, respectively.
Our spectral-only ccECP.S results in a significant improvement over the atomic spectra from previously existing tables such as STU and BFD. 
In fact, when comparing to the all-electron uncorrelated core UC, we see an improvement for the low-lying spectrum (indicated by  LMAD) as well as the full spectrum ionized down to the Ar semi-core. 
Compared to recently derived eCEPP \cite{Trail:2017jcp}, our spectral errors are only slightly larger, presumably 
due to our significantly smaller number of gaussian functions as well as shorter radial range 
of nonlocal channels.  
Since for Sc the 
higher order effects such as spin-orbit can reach 0.010~eV \cite{spin_orbit}, discrepancies below this level do not
provide a genuine quality measure 
as we have noted also above.

For the molecular calculations, we plot discrepancies from the AE CCSD(T) binding curve for a range of bond lengths. 
The compressed bond regime is plotted near the dissociation threshold, i.e., where the binding energy approaches zero while 
the vertical line indicates the equilibrium bond length. 
For ScH, we see that all ECPs result in binding discrepancies that are well within chemical accuracy with regard to the AE results, indicated by the shaded region. 
Note that STU, BFD, and UC discrepancies vary as functions of the bond length. That 
might cause some shifts  
in the predicted vibrational frequencies, whereas a flat discrepancy  should lead to better accuracy for vibrational properties. The ScO
molecule probes the charge transfer and polar
bond regime showing that both STU and BFD exhibit significantly larger deviations from the AE potential energy surface. 
Our ccECP.S results in a relatively flat discrepancy throughout the entire binding region and a very small error for the binding energy.
Due to the quality of the constructed operator, we decided that no further refinements to the ccECP.S were needed. 
The parameters for this optimal ccECP.S are given in Table \ref{tab:1st_param}.

\subsection{Titanium}\label{sec:Ti}

\begin{table*}[!htbp]
\caption{Ti AE gaps and relative errors for various core approximations. All gaps are relative to the [Ar] $3d^2 4s^2$ $^3F$ state. See Tab.\ref{tab:Sc} for further description. All in eV.}
\label{tab:Ti}
\begin{tabular}{ll|rrrrrrrr}
\hline\hline
Gaps &     &     AE &  UC & BFD &     STU &      eCEPP &   ccECP$^\dagger$\\
\hline
{[Ar] $3d^34s^2$ } &      $^4F$ &     0.020661 &  0.008558 &  0.084231 & 0.073673 &  0.005225 &  0.007623 \\
{[Ar] $3d^34s^1$}  &      $^5F$ &     0.796464 &  0.015718 &  0.110225 & 0.082066 &  0.006530 &  0.006560 \\
{[Ar] $3d^4$ }     &      $^5D$ &     3.616908 &  0.019344 &  0.185767 & 0.159201 &  0.009038 &  0.012720 \\
{[Ar] $3d^24s^1$ } &      $^4F$ &     6.804955 &  0.003393 &  0.009862 & 0.026440 & -0.003403 & -0.009247 \\
{[Ar] $3d^3$ }     &      $^4F$ &     6.915550 &  0.026609 &  0.141649 & 0.141318 &  0.020958 &  0.006538 \\
{[Ar] $3d^2$ }     &      $^3F$ &    20.387116 &  0.009622 &  0.015878 & 0.112828 &  0.008137 & -0.017891 \\
{[Ar] $3d^1$ }     &      $^2D$ &    47.867620 & -0.037232 & -0.151758 & 0.130371 & -0.040542 & -0.029817 \\
{[Ar] }            &      $^1S$ &    91.091323 & -0.124071 & -0.391450 & 0.258178 & -1.234149 &  0.012368 \\
\hline
LMAD & & &  0.014724& 0.106347& 0.09654& 0.009031& 0.008538\\
MAD & &  &   0.030568& 0.136353& 0.123009& 0.165998& 0.012846\\
\hline\hline
\end{tabular}
{\\ \small $^\dagger$ We note that our ccECP is equivalent to our ccECP.S, as described in the text. }
\end{table*}

\begin{figure*}[!htbp]
\centering
\begin{subfigure}{0.5\textwidth}
\includegraphics[width=\textwidth]{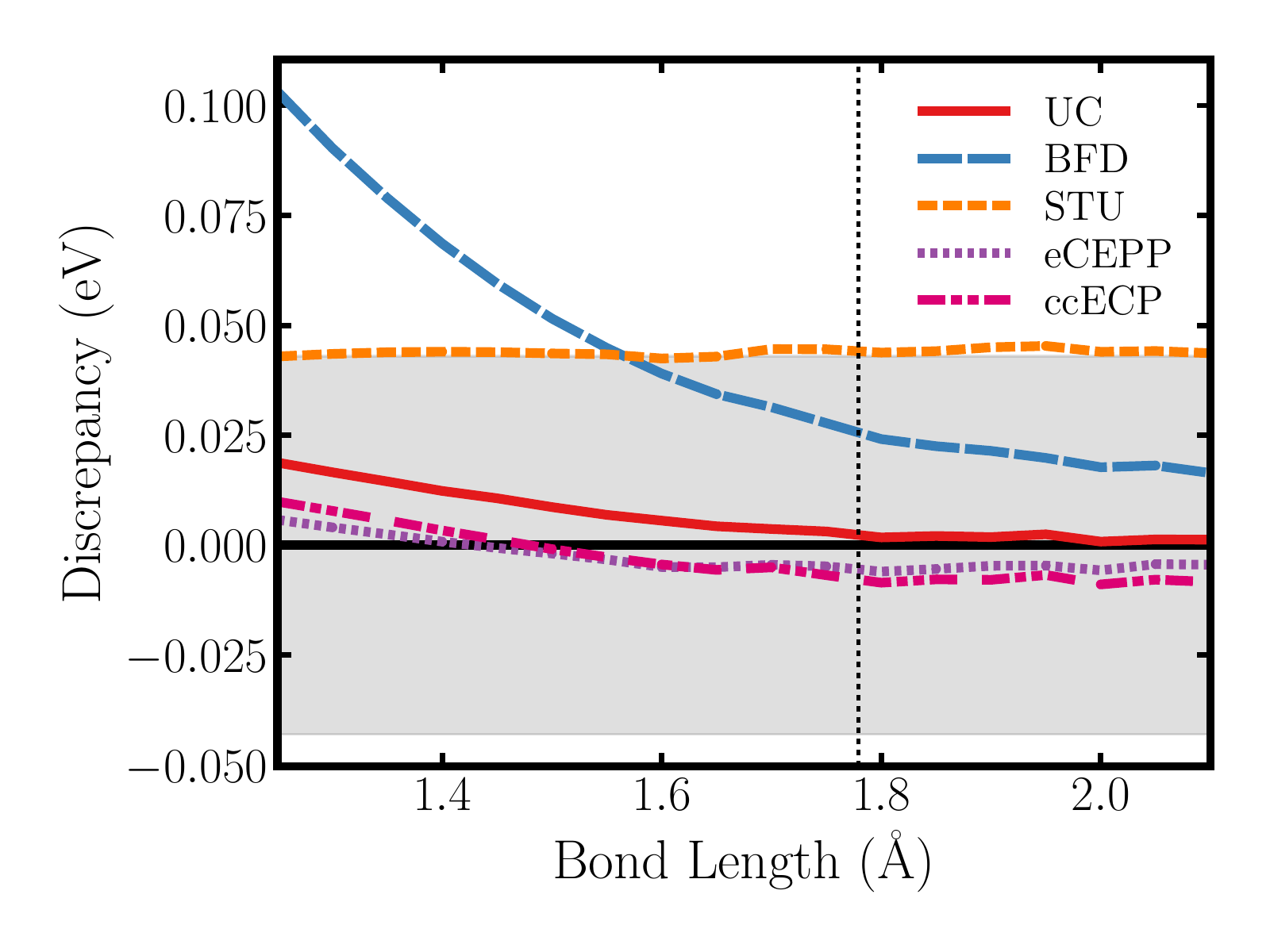}
\caption{TiH binding curve discrepancies}
\label{fig:TiH}
\end{subfigure}%
\begin{subfigure}{0.5\textwidth}
\includegraphics[width=\textwidth]{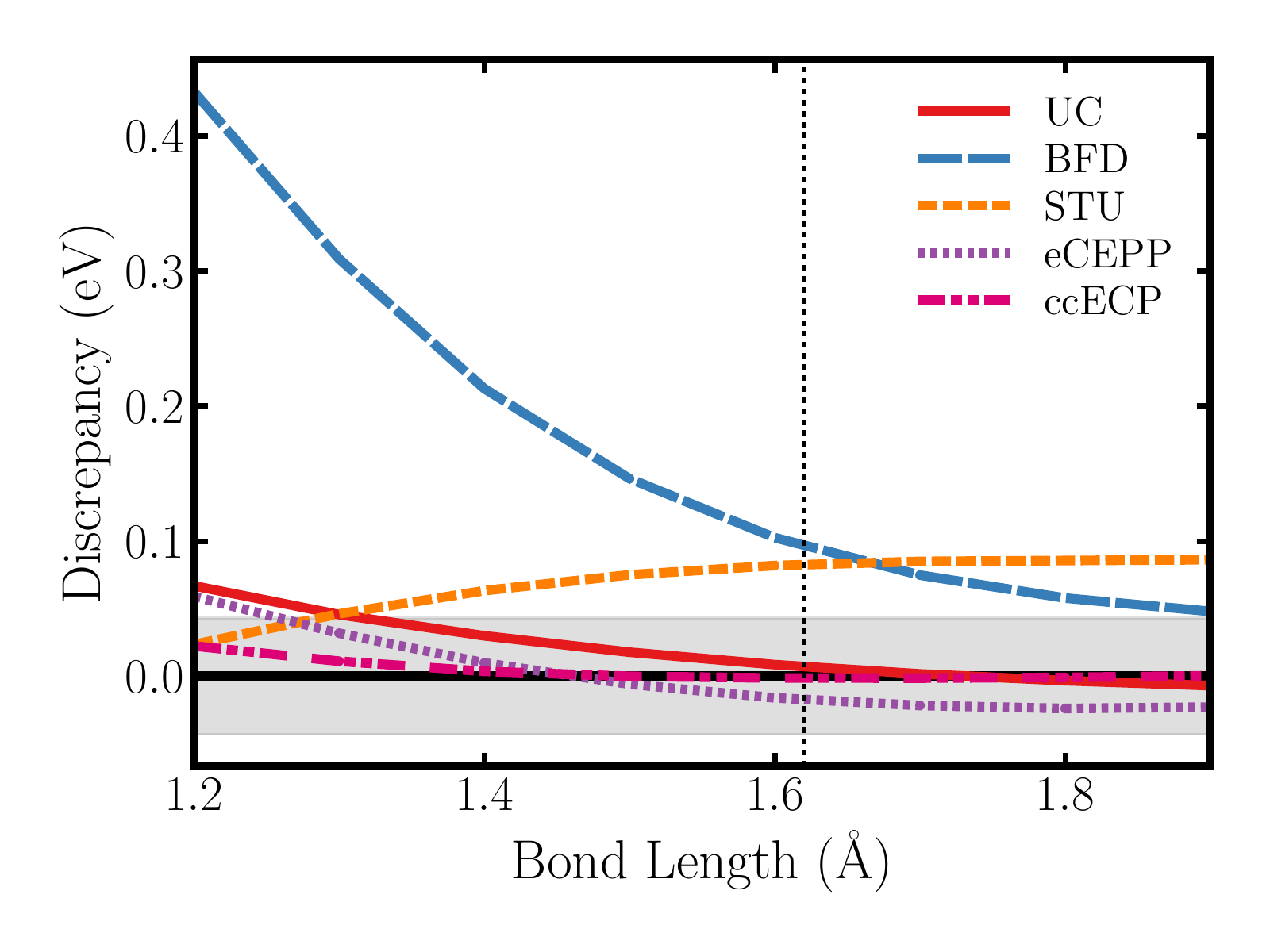}
\caption{TiO binding curve discrepancies}
\label{fig:TiO}
\end{subfigure}
\caption{Binding energy discrepancies for (a) TiH and (b) TiO molecules. The binding curves are relative to the CCSD(T) binding curve. The shaded region indicates a discrepancy of chemical accuracy in either direction.}
\label{fig:Ti_mols}
\end{figure*}

In the case of Ti, the atomic and molecular data is given in Table \ref{tab:Ti} and Fig. \ref{fig:Ti_mols}, respectively. 
For the atomic spectrum, our ccECP.S outperforms all other core approximations, for both the low lying atomic spectra (LMAD) as well as all ionizations down to the Ar semi-core. 
We see that STU and BFD result in comparable MADs of roughly 0.1~eV.
For the low-lying spectra, both the eCEPP and ccECP.S have marginally smaller discrepancies than UC showing that some of the core-valence correlations have been captured.
Note that our ccECP.S maintains uniform accuracy throughout the entire spectrum, resulting in a MAD of 0.013~eV. 

When considering the hydride molecule, the BFD shows 
significant underbinding for the compressed geometries.
On the other hand, STU maintains uniform accuracy throughout the entire binding region with mild 0.04~eV underbinding. 
UC, eCEPP, and our ccECP.S are all very comparable for TiH, showing deviations within 0.02~eV for all bond lengths. 
For TiO with polar bond, we find that BFD is inadequate to describe the binding even at equilibrium and the error reaches up to 0.4~eV near in the short bond region.
STU is also underbound near equilibrium, and has the opposite behavior to BFD in that the error decreases as the bond is compressed. 
While UC and eCEPP are well within chemical accuracy near equilibrium, each begins to underbind as the bond length is compressed. 
Near dissociation, both UC and eCEPP reach region outside the chemical accuracy. 
Our ccECP.S, on the other hand, is well within the desired error margin throughout the entire bonding region. 
Additionally, the ccECP.S has an extremely flat discrepancy near the equilibrium.  
Considering the accuracy of the constructed ccECP.S, we did not pursue any refinements. 
The parameters for the Ti ccECP.S are given in Table \ref{tab:1st_param}.

\subsection{Vanadium}\label{sec:V}

\begin{table*}[!htbp]
\centering
\caption{V AE gaps and relative errors for various core approximations. All gaps are relative to the [Ar] $3d^3 4s^2$ $^4F$ state. See Tab.\ref{tab:Sc} for further description. All data in eV.}
\label{tab:V}
\begin{tabular}{ll|rrrrrrrrr}
\hline\hline
Gaps &    &       AE &        UC&       BFD &     STU &      eCEPP & ccECP.S & ccECP\\
\hline

{[Ar] $3d^4 4s^2$ } &      $^5D$ &    -0.476250 &  0.010698 &  0.104690 & 0.069543 & -0.002350 &  0.005158 &  0.018813 \\
{[Ar] $3d^4 4s^1$}  &      $^6D$ &     0.225411 &  0.016419 &  0.122574 & 0.072073 & -0.012547 &  0.000018 &  0.016268 \\
{[Ar] $3d^5$}       &      $^6S$ &     2.477803 &  0.022536 &  0.223108 & 0.153314 & -0.012134 &  0.009320 &  0.040209 \\
{[Ar] $3d^3 4s^1$}  &      $^5F$ &     7.040745 &  0.003875 &  0.011638 & 0.014400 & -0.012391 & -0.012788 & -0.005745 \\
{[Ar] $3d^4$}       &      $^5D$ &     6.702435 &  0.028646 &  0.161530 & 0.123399 &  0.000007 &  0.003149 &  0.030633 \\
{[Ar] $3d^3$}       &      $^4F$ &    21.353719 &  0.012672 &  0.037204 & 0.086384 &  0.004329 & -0.009391 &  0.011301 \\
{[Ar] $3d^2$}       &     $^3 F$ &    50.673774 & -0.030859 & -0.104118 & 0.088290 & -0.015417 & -0.022190 & -0.008337 \\
{[Ar] $3d^1$}       &      $^2D$ &    97.385632 & -0.106347 & -0.297488 & 0.181266 & -0.065904 & -0.024749 & -0.014226 \\
{[Ar]}              &      $^1S$ &   162.607898 & -0.225926 & -0.639514 & 0.434473 & -1.746391 &  0.012854 &  0.026267 \\
\hline
LMAD & & & 0.016435& 0.124708& 0.086546& 0.007886& 0.006087& 0.022334\\ 
MAD & & &  0.050886& 0.189096& 0.135905& 0.207941& 0.011069& 0.019089\\
\hline\hline
\end{tabular}
\end{table*}

\begin{figure*}[!htbp]
\centering
\begin{subfigure}{0.5\textwidth}
\includegraphics[width=\textwidth]{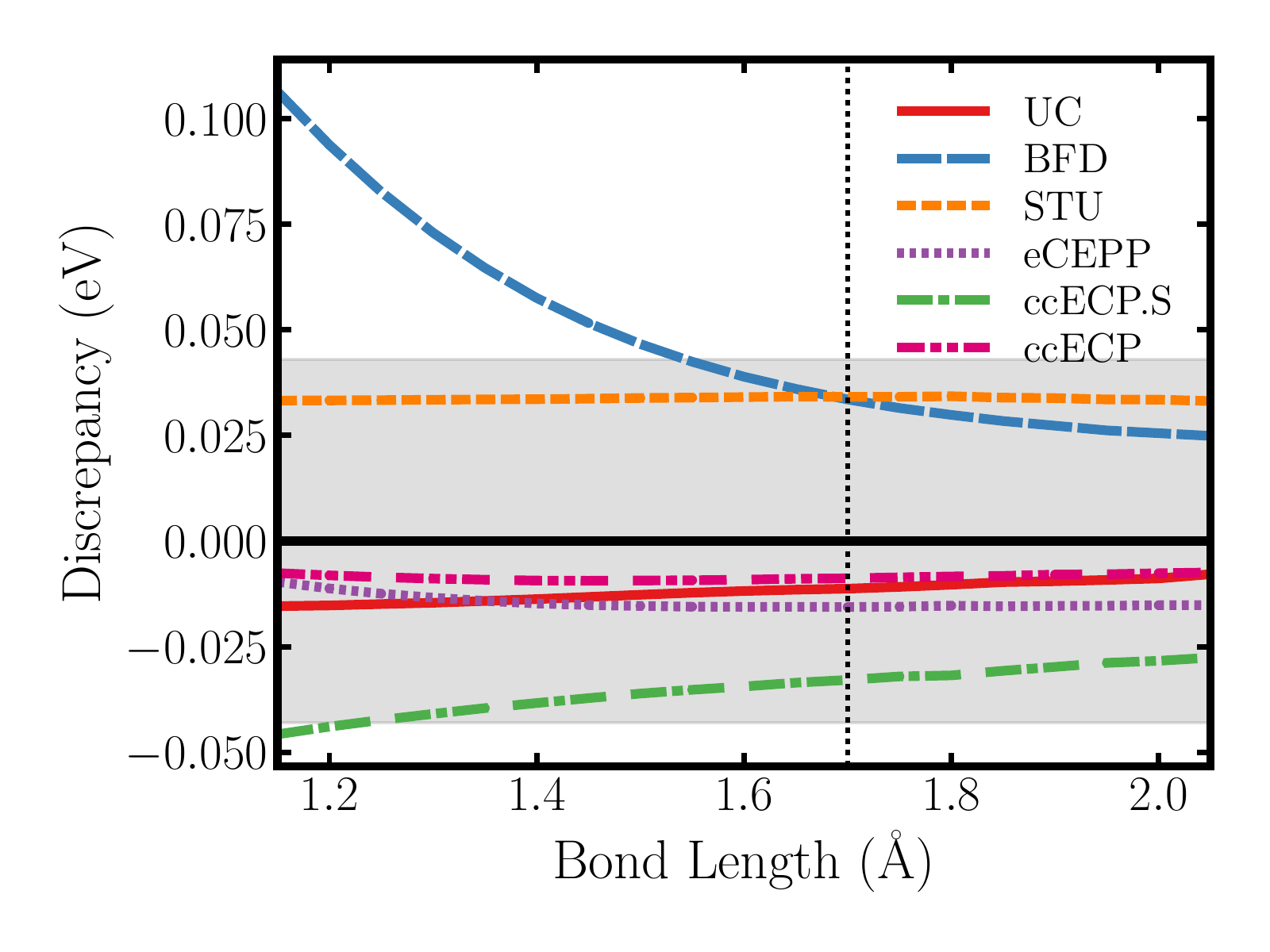}
\caption{VH binding curve discrepancies.}
\label{fig:VH}
\end{subfigure}%
\begin{subfigure}{0.5\textwidth}
\includegraphics[width=\textwidth]{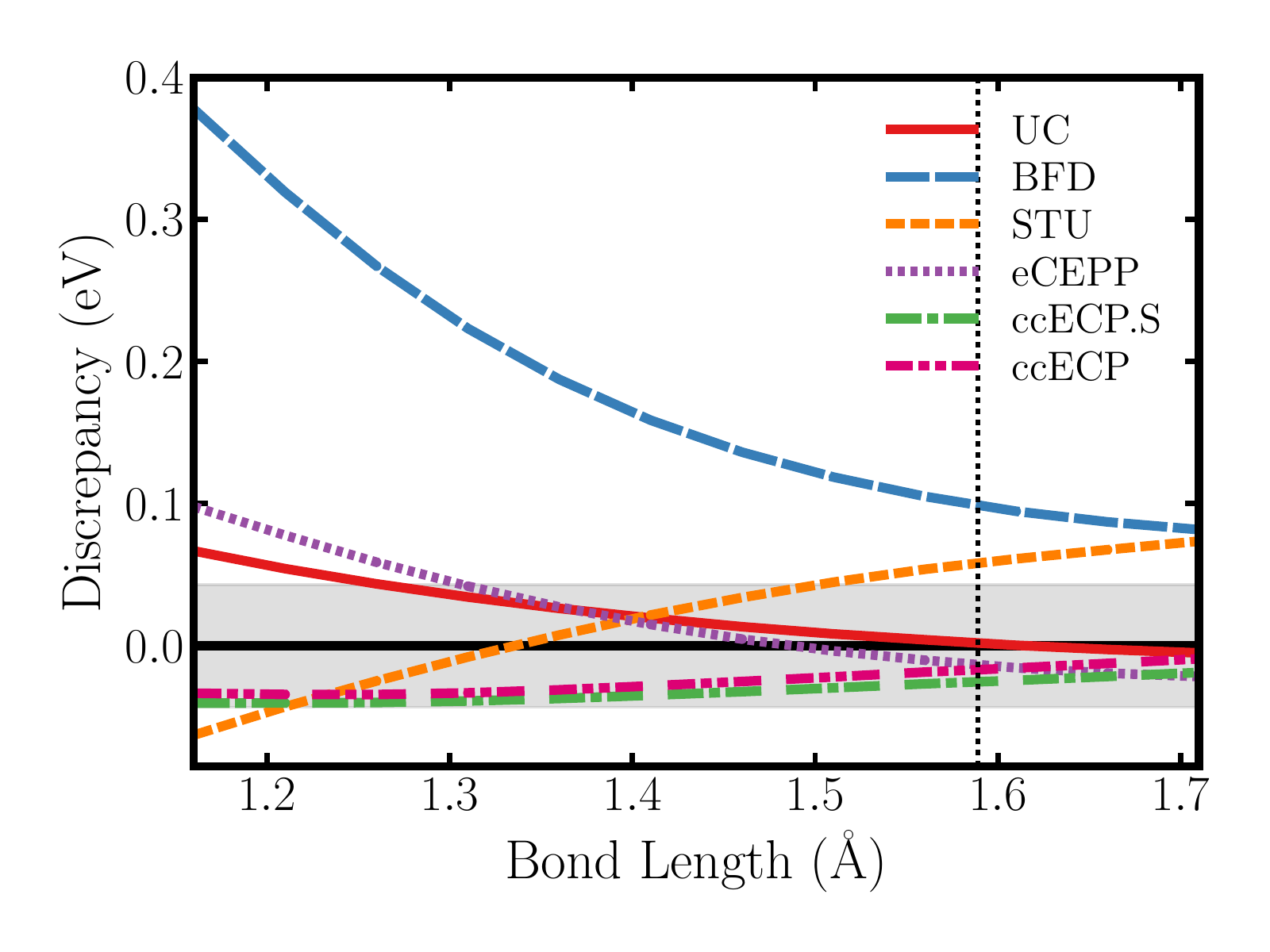}
\caption{VO binding curve discrepancies}
\label{fig:VO}
\end{subfigure}
\caption{Binding energy discrepancies for (a) VH and (b) VO molecules. The binding curves are relative to the CCSD(T) binding curve. The shaded region indicates a discrepancy of chemical accuracy in either direction.}
\label{fig:V_mols}
\end{figure*}

The atomic and molecular data for the V ECPs is given in Table \ref{tab:V} and Fig. \ref{fig:V_mols}, respectively. 
For the atomic spectra, our ccECP.S has significantly smaller discrepancies compared to all other core approximations, including the all-electron UC approximation, for both the low-lying spectra and all ionizations. 
If we consider VO for our ccECP.S, we see that we have quite favorable binding properties compared to other core approximations. 
Both BFD and STU exhibit relatively large errors throughout the entire bonding region, including near equilibrium. 
Both eCEPP and UC begin to underbind to nearly 0.1~eV near the dissociation threshold, whereas our ccECP.S is well within chemical accuracy. 
However, when considering the hydride, we found that our spectral-only ECP overbinds by  a small amount of $\approx$ 0.04 eV
that reaches the borderline at dissociation. In fact, this is still acceptable considering 
the systematic biases present although less accurate than the eCEPP binding curve. 
Note that ccECP.S ECPs fits a significant part of
 the atomic a spectrum that includes ionizations past the valence electrons and down into the 3$s$ and 3$p$ semi-core states. However, this nudges
 the ECP operator away from 
 transferability optimum as can be observed on the VH molecule.
Therefore, we refined ccECP.S by reducing the considered spectrum and modifying our objective function as described in part \ref{Objective}, resulting in refined ccECP. 
The charge-transfer physics is unchanged 
and it is almost identical to the spectral ccECP.S. 
However, we see a significant improvement for the hydride molecule.
In terms of the spectrum, we slightly compromise the low-lying spectrum (LMAD) while maintaining roughly the same overall MAD. 
The parameters for our ccECP of V are included in Table \ref{tab:1st_param}.

\subsection{Chromium}\label{sec:Cr}

\begin{figure*}[!htbp]
\centering
\begin{subfigure}{0.5\textwidth}
\includegraphics[width=\textwidth]{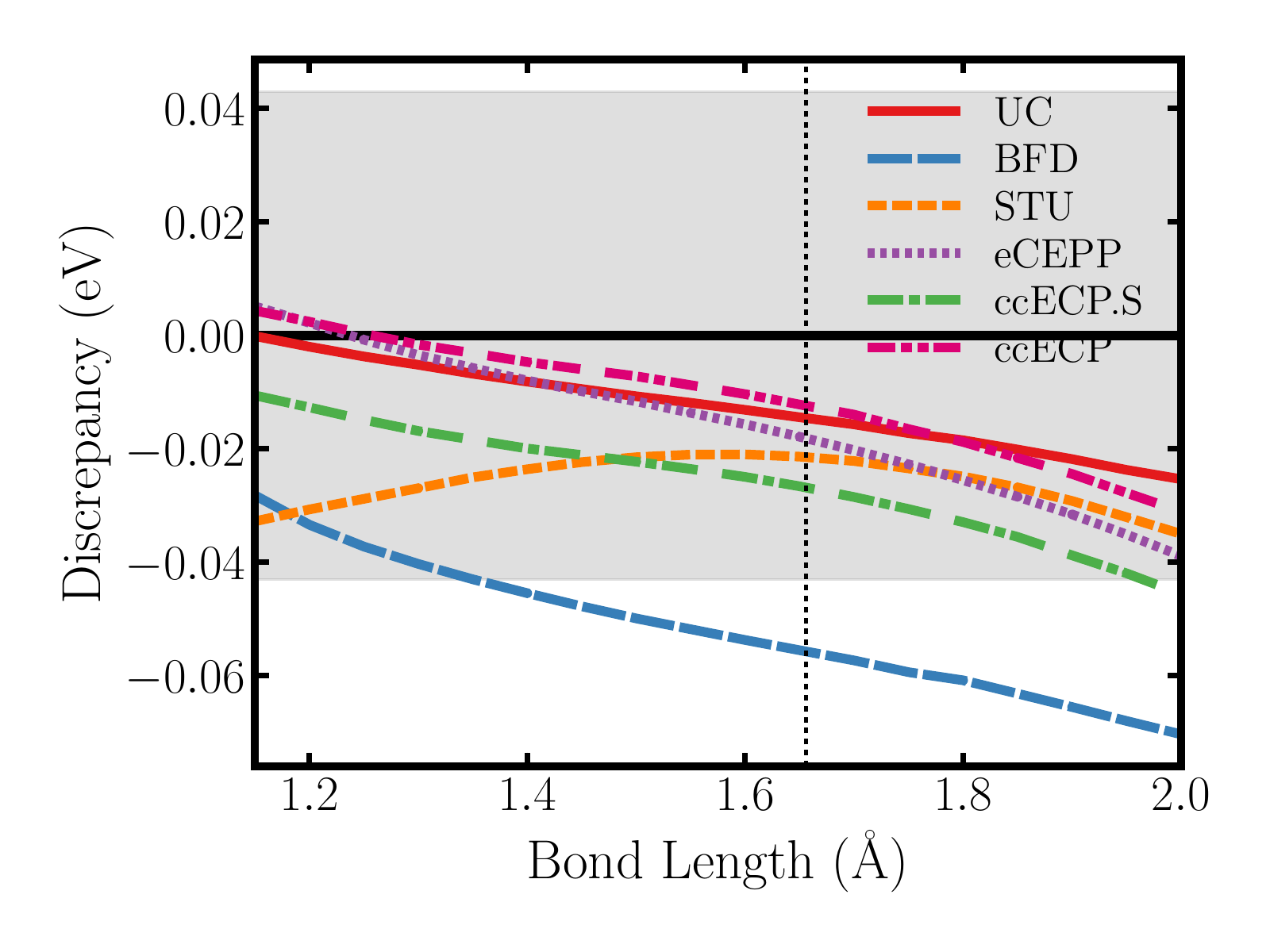}
\caption{CrH binding curve discrepancies.}
\label{fig:CrH}
\end{subfigure}%
\begin{subfigure}{0.5\textwidth}
\includegraphics[width=\textwidth]{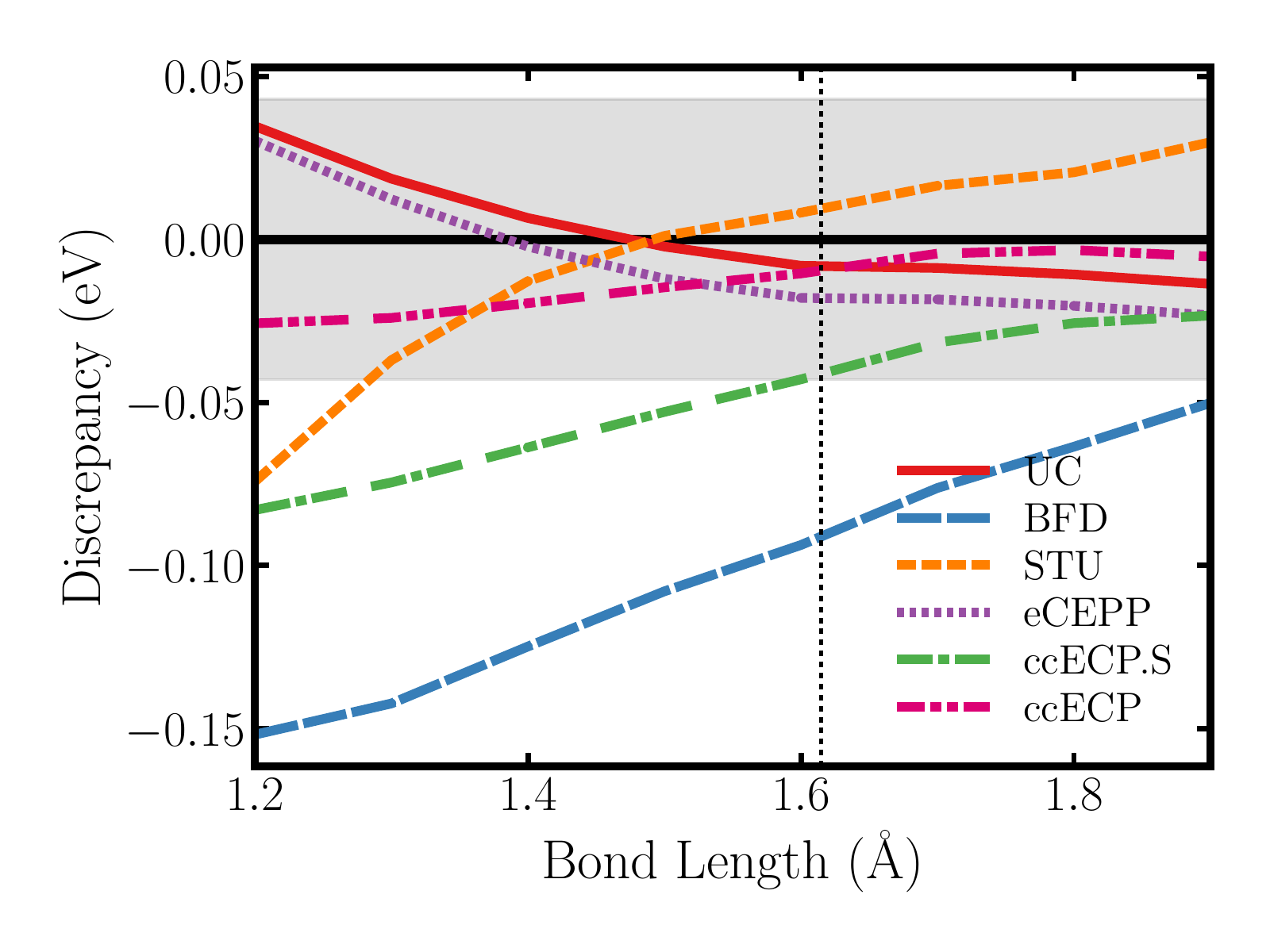}
\caption{CrO binding curve discrepancies}
\label{fig:CrO}
\end{subfigure}
\caption{Binding energy discrepancies for (a) CrH and (b) CrO molecules. The binding curves are relative to the CCSD(T) binding curve. The shaded region indicates a discrepancy of chemical accuracy in either direction.}
\label{fig:Cr_mols}
\end{figure*}

\begin{table*}[!htbp]
\centering
\caption{Cr AE gaps and relative errors for various ECPs. All gaps are relative to the [Ar] $3d^4 4s^2$ $^5D$ state. See Tab.\ref{tab:Sc} for further description. All values in eV.}
\label{tab:Cr}
\begin{tabular}{ll|rrrrrrrrr}
\hline\hline
Gaps &   &        AE &   UC &     BFD &      STU &   eCEPP &  ccECP.S & ccECP \\
\hline
{[Ar] $3d^54s^2$  } &      $^6S$ &    -1.639160 &  0.008708 &  0.100464 & 0.075539 &  0.002912 &  0.002231 & -0.007320 \\
{[Ar] $3d^54s^1$ }  &      $^7S$ &    -1.026768 &  0.012626 &  0.109308 & 0.070858 & -0.002395 & -0.001850 & -0.015864 \\
{[Ar] $3d^6$ }      &      $^5D$ &     3.444963 &  0.006449 &  0.189582 & 0.145908 & -0.006204 & -0.002068 & -0.024681 \\
{[Ar] $3d^44s^1$ }  &      $^6D$ &     7.256491 &  0.003483 &  0.003102 & 0.016517 & -0.016762 & -0.011701 & -0.017497 \\
{[Ar] $3d^5$ }      &      $^6S$ &     5.735591 &  0.023783 &  0.139894 & 0.140656 &  0.010259 &  0.016681 & -0.005632 \\
{[Ar] $3d^4$ }      &      $^5D$ &    22.265020 &  0.013198 &  0.032490 & 0.127077 & -0.010531 &  0.006177 & -0.003946 \\
{[Ar] $3d^3$ }      &      $^4F$ &    53.342070 & -0.022558 & -0.082097 & 0.181337 & -0.015293 & -0.003129 &  0.008626 \\
{[Ar] $3d^2$ }      &      $^3F$ &   102.464577 & -0.084655 & -0.217174 & 0.344850 &  0.022232 & -0.012436 &  0.032518 \\
{[Ar] $3d^1$ }      &      $^2D$ &   171.920939 & -0.176820 & -0.446866 & 0.669754 &  0.143731 & -0.017361 &  0.082695 \\
{[Ar]}              &      $^1S$ &   262.444038 & -0.312659 & -1.402775 & 1.242200 &  0.422348 &  0.020817 &  0.217202 \\
\hline
LMAD & & & 0.011010& 0.10847& 0.089896& 0.007706& 0.006906& 0.014199\\
MAD & & &  0.066494& 0.272375& 0.30147& 0.065267& 0.009445& 0.041598\\
\hline\hline
\end{tabular}
\end{table*}
The atomic and molecular data for Cr are given in Table \ref{tab:Cr} and Fig. \ref{fig:Cr_mols}, respectively. 
For consistency with the rest of the transition metals, we choose to show excitation energies with respect to [Ar] $3d^4 4s^2$ $^5D$ state instead of the ground state which is [Ar] $3d^5 4s^1$ $^7S$. 
Our spectral ECP, ccECP.S, has a significantly improved spectrum when compared to all other core approximations, including the all-electron UC approximation.  
The low-lying spectrum (LMAD) is slightly improved over eCEPP, while the MAD across the spectrum above [Ar] is significantly lower. 
The ccECP.S performs reasonably well for CrH, being within chemical accuracy to AE for the entire binding region. 
However, when we consider the oxide CrO, we see that the ccECP.S shows clear tendency to overbind, up to $\approx 0.08$~eV near the dissociation threshold. 
Our refined ccECP, optimized using Eq. \ref{eqn:final_objective}, shows    well balanced atomic spectrum
with mild increase in LMAD. In molecular calculations, 
 we see a modest improvement for CrH and a very satisfactory reduction in overbinding when compared to ccECP.S.
The final ccECP is well within chemical accuracy throughout the entire binding region, including near the dissociation threshold. 
The parameters for our ccECP are given in Table \ref{tab:1st_param}.

\subsection{Manganese}\label{sec:Mn}

\begin{table*}[!htbp]
\centering
\caption{Mn AE gaps and relative error for various core approximations. All gaps are relative to the [Ar] $3d^5 4s^2$ $^6S$ state. See Tab.\ref{tab:Sc} for further description. All in eV.}
\label{tab:Mn}
\begin{tabular}{ll|rrrrrrrrr}
\hline\hline
Gaps &     &      AE &    UC  &    BFD &     STU &      eCEPP & ccECP.S & ccECP \\
\hline
{[Ar] $3d^64s^2$  }  & $^5D$&        1.299360 &  0.001050 &  0.146607 &  0.028564 &  0.004187 &  -0.003555 &   -0.006144 \\
{[Ar] $3d^64s^1$ }   & $^6D$&        2.151429 &  0.006575 &  0.169347 &  0.010605 &  0.002082 &  -0.006991 &   -0.013559 \\
{[Ar] $3d^7$ }       & $^4F$&        5.772304 &  0.007078 &  0.285947 &  0.059634 &  0.018798 &   0.004177 &   -0.003478 \\
{[Ar] $3d^54s^1$ }   & $^7S$&        7.414670 &  0.003239 & -0.003639 & -0.004177 & -0.006771 &  -0.008937 &   -0.008985 \\
{[Ar] $3d^6$ }       & $^5D$&        9.245516 &  0.007056 &  0.194099 &  0.017318 &  0.003874 &  -0.015398 &   -0.013276 \\
{[Ar] $3d^5$ }       & $^6S$&       23.059159 &  0.014572 &  0.022398 &  0.081569 &  0.014290 &   0.030141 &    0.050613 \\
{[Ar] $3d^4$ }       & $^5D$&       56.822155 & -0.015390 & -0.072555 &  0.256077 & -0.026524 &   0.016815 &    0.033293 \\
{[Ar] $3d^3$ }       & $^4F$&      108.315370 & -0.074830 & -0.158482 &  0.600224 & -0.059576 &   0.010784 &    0.017702 \\
{[Ar] $3d^2$ }       & $^3F$&      180.815120 & -0.162619 & -0.291000 &  1.159615 & -0.070647 &   0.003077 &    0.004779 \\
{[Ar] $3d^1$ }       & $^2D$&      276.392004 & -0.282966 & -0.586852 &  1.989336 & -0.043384 &  -0.008268 &    0.013606 \\
{[Ar]}               & $^1S$&      395.519402 & -0.451819 & -1.184772 &  3.166735 &  0.029970 &   0.001807 &    0.103825 \\
\hline
LMAD & & &  0.005000& 0.159928& 0.02406& 0.007142& 0.007812& 0.009088\\
MAD & & &   0.093381& 0.283245& 0.67035& 0.025464& 0.009995& 0.024478\\
\hline\hline
\end{tabular}
\end{table*}

\begin{figure*}[!htbp]
\centering
\begin{subfigure}{0.5\textwidth}
\includegraphics[width=\textwidth]{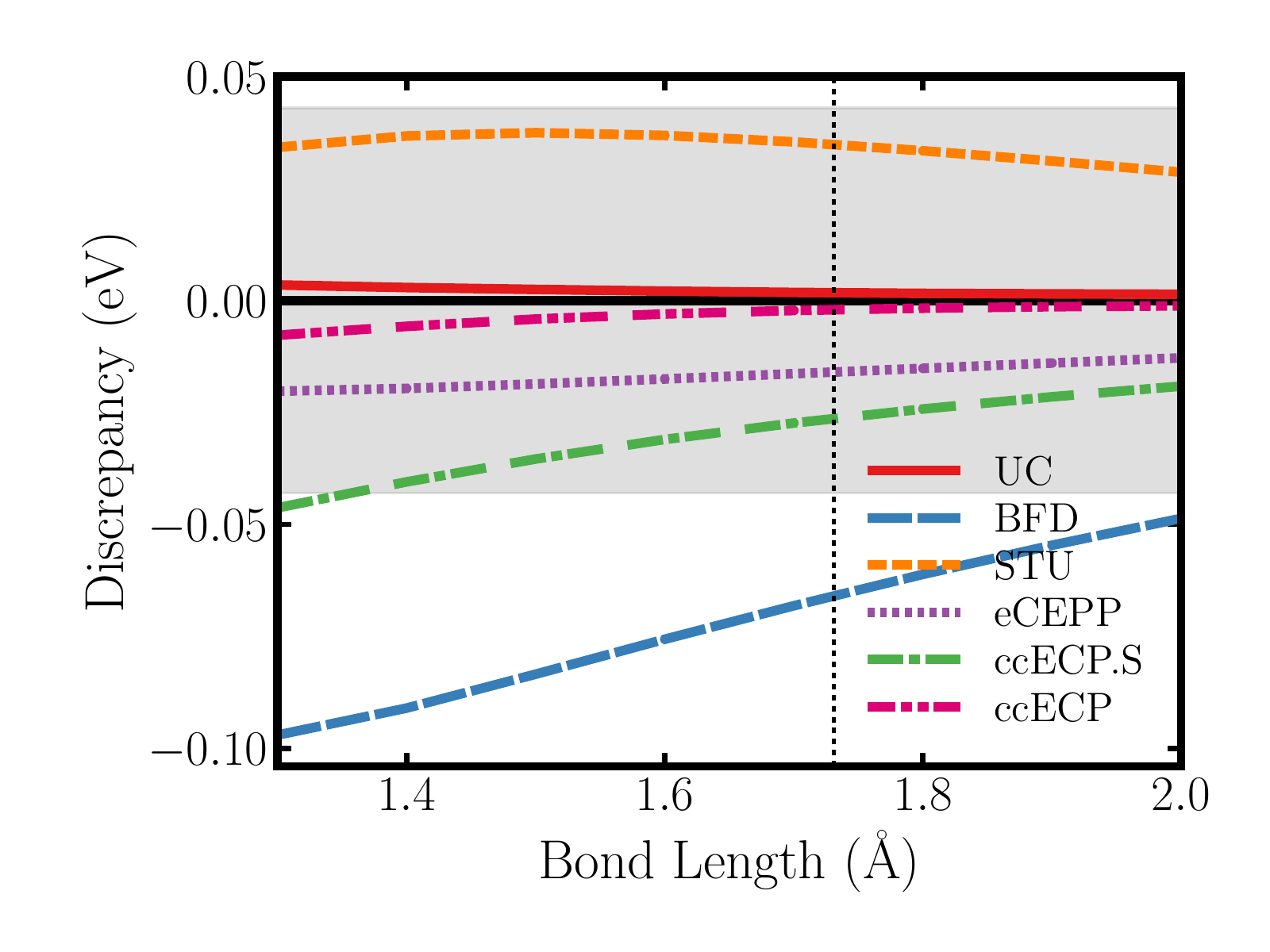}
\caption{MnH binding curve discrepancies.}
\label{fig:MnH}
\end{subfigure}%
\begin{subfigure}{0.5\textwidth}
\includegraphics[width=\textwidth]{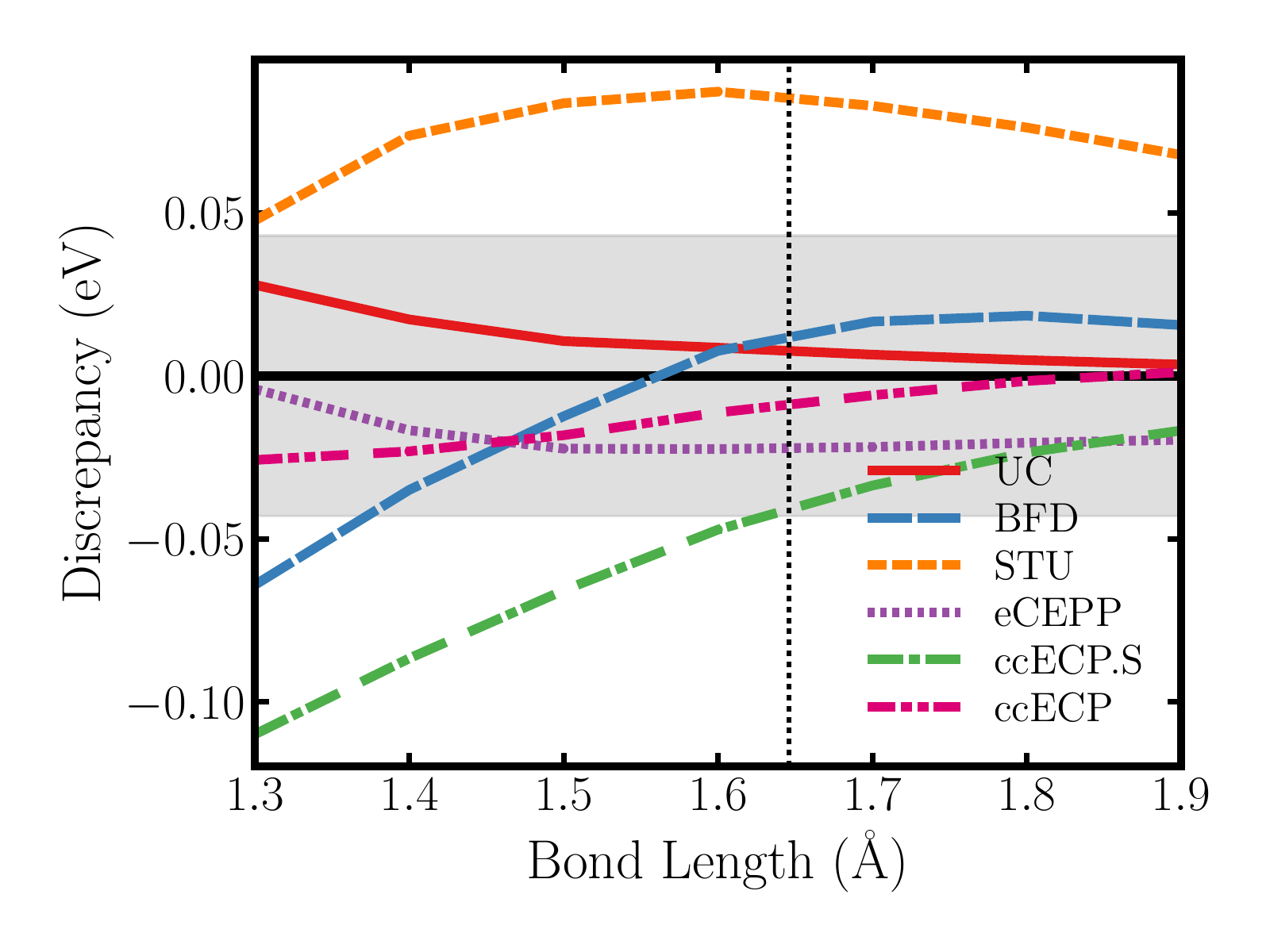}
\caption{MnO binding curve discrepancies}
\label{fig:MnO}
\end{subfigure}
\caption{Binding energy discrepancies for (a) MnH and (b) MnO molecules. The binding curves are relative to the CCSD(T) binding curve. The shaded region indicates a discrepancy of chemical accuracy in either direction.}
\label{fig:Mn_mols}
\end{figure*}

The atomic and molecular results for Mn are given in Table \ref{tab:Mn} and Figure \ref{fig:Mn_mols}, respectively. 
For the spectral ccECP.S, we find similar uniform accuracy for both the low-lying atomic and full  spectrum to [Ar].  
It is interesting to see that despite this 
fidelity we find the overall transferability to be lacking. 
While ccECP.S for MnH is within chemical accuracy for most of the binding curve we observe a slightly larger discrepancy near the dissociation threshold. 
This error is more severe for MnO with overbinding by roughly 0.04~eV near equilibrium that increases to 0.11~eV at dissociation. 
Therefore, despite having the best atomic spectrum, the ccECP.S MnO molecule has the largest discrepancies from AE when compared to all other core approximations. 
We clearly observe that within the use parameterization, the atomic accuracy and molecular accuracy pull in rather opposite directions. 

We improve upon our spectral only ccECP.S and reoptimize via Eq. \ref{eqn:final_objective} to obtain our final ccECP. 
In terms of the atomic spectrum, we slightly increase the LMAD to $\approx$ 0.01~eV. 
Some error of 0.1~eV is introduced to the highly ionized [Ar] semi-core excitation.
This increase results in a MAD on the entire spectrum of only 0.024~eV, which outperforms all other core approximations including the all-electron UC. 
For both the hydride and oxide we see a clear improvement.
For MnH, the ccECP has below a 0.01~eV discrepancy for all bond lengths.
The most drastic improvement comes for the oxide molecule, where the discrepancy is bounded by the chemical accuracy throughout the entire curve.
The resulting ccECP has excellent atomic and molecular properties  and the parameters are given in Table \ref{tab:1st_param}.

\subsection{Iron}\label{sec:Fe}

\begin{table*}[!htbp]
\centering
\caption{Fe AE gaps and relative errors for various ECPs. All gaps are relative to the [Ar] $3d^6 4s^2$ $^5D$ state. See Tab.\ref{tab:Sc} for further description. All data in eV.}
\label{tab:Fe}
\begin{tabular}{ll|rrrrrrrrrrrrr}
\hline\hline

Gaps & &           AE &        UC &       BFD &      STU &     eCEPP &   ccECP.S &     ccECP \\
\hline
{[Ar]$4s^23d^7$ }  &      $^4F$ &    -0.057987 &  0.001660 &  0.156504 & 0.044160 &  0.011820 &  0.007696 & -0.005231 \\
{[Ar] $4s^1 3d^7$} &      $^5F$ &     0.888551 &  0.006370 &  0.171349 & 0.033548 &  0.008323 & -0.002355 & -0.014157 \\
{[Ar] $3d^8$}      &      $^3F$ &     4.157110 &  0.010139 &  0.291230 & 0.092898 &  0.030171 & -0.003989 & -0.017050 \\
{[Ar] $4s^1 3d^6$} &      $^6D$ &     7.885826 &  0.004183 &  0.001180 & 0.005114 & -0.004736 & -0.020115 & -0.015930 \\
{[Ar] $3d^7$}      &      $^4F$ &     8.162522 &  0.016486 &  0.199663 & 0.071353 &  0.026536 &  0.002202 & -0.002154 \\
{[Ar]$3d^6$ }      &      $^5D$ &    24.090858 &  0.014936 &  0.033942 & 0.099740 &  0.010690 &  0.008389 &  0.013777 \\
{[Ar]$3d^5$ }      &      $^6S$ &    54.661521 & -0.005615 & -0.168899 & 0.206248 & -0.017071 &  0.004946 &  0.021878 \\
{[Ar]$3d^4$}       &      $^5D$ &   109.577638 & -0.051901 & -0.192542 & 0.455601 & -0.007441 & -0.005419 & -0.008571 \\
{[Ar]$3d^3$  }     &      $^4F$ &   185.022592 & -0.129771 & -0.222149 & 0.890969 &  0.067005 & -0.000924 & -0.005565 \\
{[Ar]$3d^2$ }      &      $^3F$ &   284.232562 & -0.235873 & -0.354853 & 1.557802 &  0.247372 &  0.001835 &  0.040188 \\
{[Ar]$3d^1$ }      &      $^2D$ &   409.151353 & -0.374283 & -0.745345 & 2.506921 &  0.518200 & -0.005224 &  0.160413 \\
{[Ar] }            &      $^1S$ &   560.024014 & -0.563164 & -1.561520 & 3.821096 &  0.924669 &  0.002865 &  0.431303 \\
\hline
LMAD & & &  0.007768& 0.163985& 0.049415& 0.016317& 0.007271& 0.010904\\
MAD & & &   0.117865& 0.341598& 0.815454& 0.156170& 0.005497& 0.061351\\
\hline\hline
\end{tabular}
\end{table*}

\begin{figure*}[!htbp]
\centering
\begin{subfigure}{0.5\textwidth}
\includegraphics[width=\textwidth]{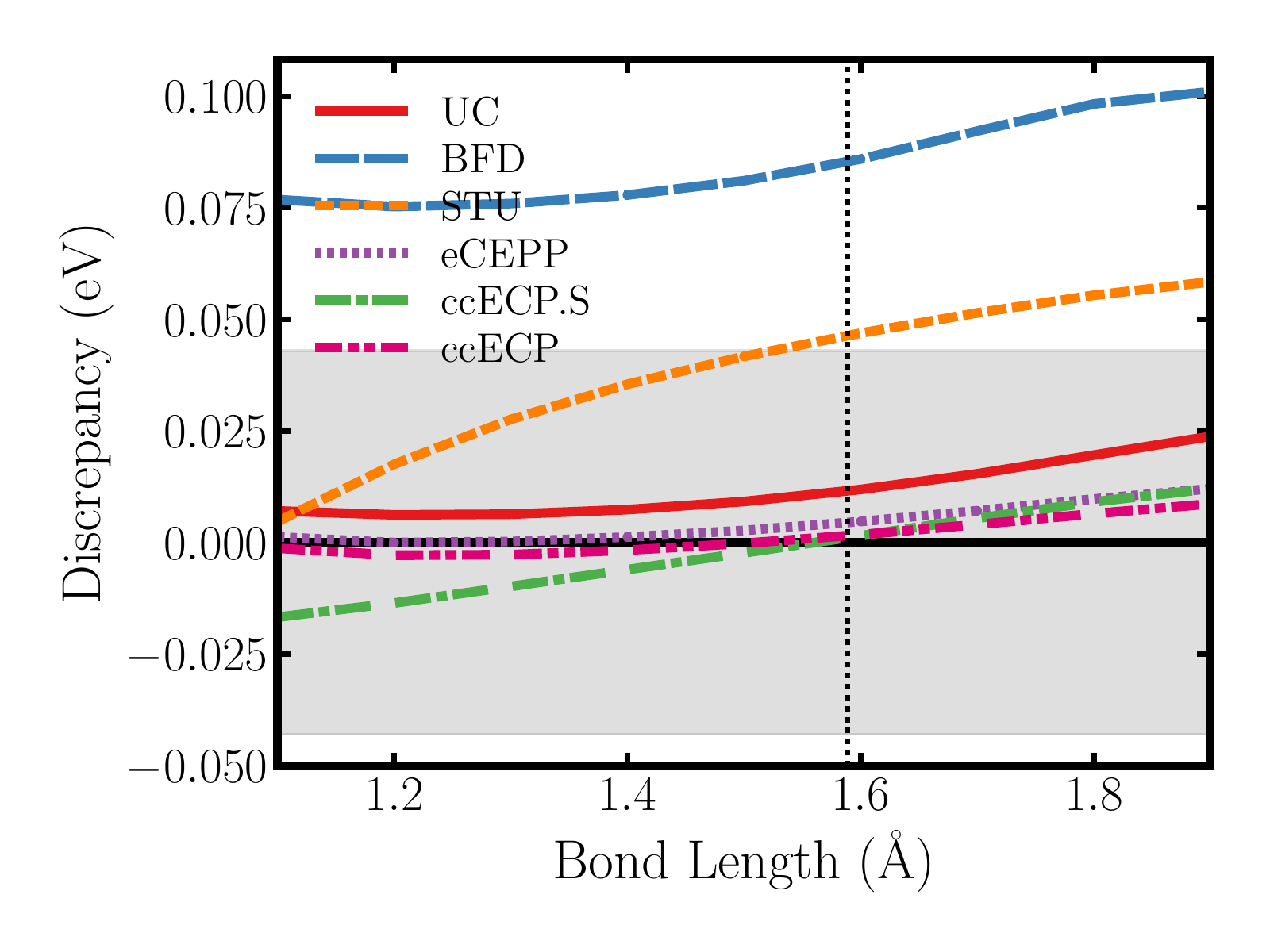}
\caption{FeH binding curve discrepancies.}
\label{fig:FeH}
\end{subfigure}%
\begin{subfigure}{0.5\textwidth}
\includegraphics[width=\textwidth]{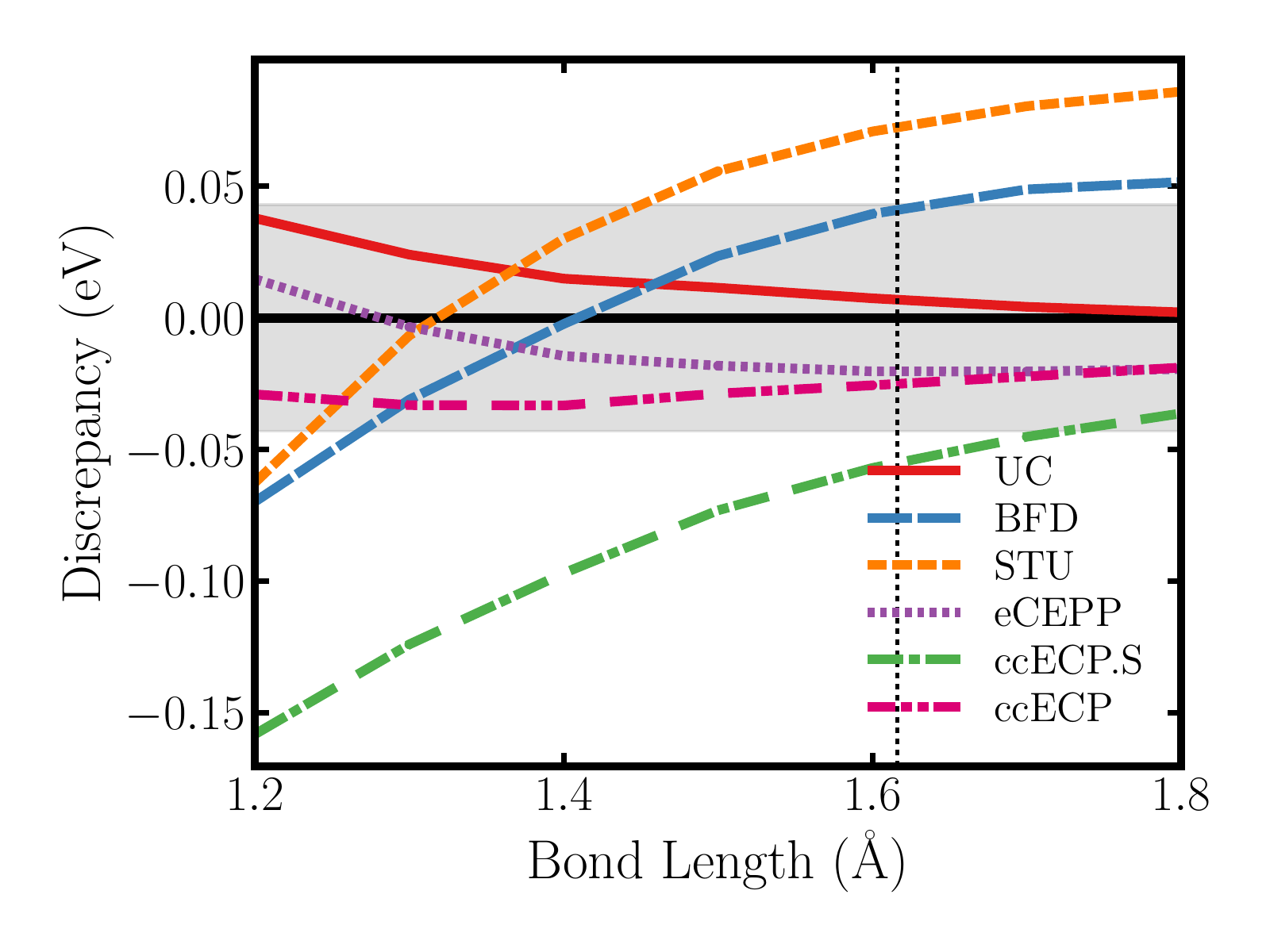}
\caption{FeO binding curve discrepancies}
\label{fig:FeO}
\end{subfigure}
\caption{Binding energy discrepancies for (a) FeH and (b) FeO molecules. The binding curves are relative to the CCSD(T) binding curve. The shaded region indicates a discrepancy of chemical accuracy in either direction.}
\label{fig:Fe_mols}
\end{figure*}

The atomic discrepancies for Fe atom are given in Table \ref{tab:Fe}.
Our spectral ccECP.S produces an atomic spectrum with uniform accuracy throughout the entire valence spectrum, with a low-lying spectrum MAD of 7~meV, and 5~meV for the entire spectrum.
In terms of the low-lying spectrum (LMAD), the ccECP.S is comparable to the all-electron UC approximation. In addition,
the ccECP.S has a much lower MAD for the entire spectrum. 
While the ccECP.S is well within chemical accuracy for FeH, when we consider the polar bond, illustrated by the FeO molecule, we find significant overbinding across the entire binding region (Figure \ref{fig:Fe_mols}).
In fact, the ccECP.S overbinds well outside chemical accuracy for most of the bond lengths. 
It should be noted that complications related to molecules with high-spin TM atoms 
are clearly visible  
for BFD and STU ECPs 
with significant discrepancies in the binding energies as well as in the shapes of the potential energy surfaces. 

We achieve a remarkable balance between the atomic spectrum as well as the molecular properties with our refined ccECP. 
The ccECP results in an increase of only 3~meV on the LMAD, and we find larger errors only for the highest ionizations. 
It should be noted that the absolute error of 0.43~eV for the ionization to the [Ar] semi-core corresponds to relative error of only $\approx$ 0.08 \% since the excitation is sizeable, 560~eV. 
Additionally, despite the compromise in the spectrum, the ccECP outperforms all other ECPs on the LMAD and all the core approximations on the entire MAD.
The molecular properties are significantly improved for the ccECP.
For FeH, there is near perfect agreement on the binding curve, both near equilibrium as well as near dissociation. 
For FeO, we see a significant reduction in the overbinding, putting the binding discrepancy within the  chemical accuracy of the AE CCSD(T) binding curve. 
The parameters for this ccECP are given in Table \ref{tab:2nd_param}. 

\subsection{Cobalt}\label{sec:Co}

\begin{table*}[!htbp]
\centering
\caption{Co AE gaps and relative errors for various ECPs. All gaps are relative to the [Ar] $3d^7 4s^2$ $^4F$ state. See Tab.\ref{tab:Sc} for further description. All data in eV.}
\label{tab:Co}
\begin{tabular}{ll|rrrrrrrr}
\hline\hline
Gaps &    &       AE &   UC  &    BFD &     STU &  ccECP.S & ccECP\\
\hline
{[Ar] $3d^84s^2$}  & $^3F$ &     -0.648824 &  0.003846 &  0.138580 &  0.038570 &    0.003500 &         -0.011140 \\
{[Ar] $3d^84s^1$ } & $^4F$ &      0.404998 &  0.008382 &  0.146827 &  0.029158 &    0.000131 &         -0.018138 \\
{[Ar] $3d^9$ }     & $^2D$ &      3.290224 &  0.013495 &  0.245172 &  0.082758 &    0.001712 &         -0.027850 \\
{[Ar] $3d^74s^1$ } & $^5F$ &      8.285103 &  0.004751 & -0.004194 &  0.008731 &   -0.007956 &         -0.014990 \\
{[Ar] $3d^8$ }     & $^3F$ &      7.852169 &  0.016852 &  0.154921 &  0.060521 &    0.007556 &         -0.012459 \\
{[Ar] $3d^7$ }     & $^4F$ &     24.959596 &  0.014340 &  0.012356 &  0.096283 &    0.010968 &          0.005804 \\
{[Ar] $3d^6$ }     & $^5D$ &     58.474931 & -0.000315 & -0.137538 &  0.215407 &    0.000966 &          0.010511 \\
{[Ar] $3d^5$ }     & $^6S$ &    109.967312 & -0.035818 & -0.288764 &  0.449440 &   -0.020260 &          0.016728 \\
{[Ar] $3d^4$ }     & $^5D$ &    189.676472 & -0.096964 & -0.229092 &  0.878645 &   -0.018478 &          0.002135 \\
{[Ar] $3d^3$ }     & $^4F$ &    292.523396 & -0.190846 & -0.217323 &  1.539629 &    0.001712 &          0.062171 \\
{[Ar] $3d^2$ }     & $^3F$ &    421.774512 & -0.312948 & -0.383028 &  2.475055 &    0.017756 &          0.218327 \\
{[Ar] $3d^1$ }     & $^2D$ &    579.286933 & -0.467425 & -0.905317 &  3.738709 &    0.010326 &          0.520393 \\
{[Ar] }            & $^1S$ &    765.100530 & -0.673798 & -1.957082 &  5.417765 &   -0.013019 &          1.067807 \\
\hline 
LMAD & & & 0.009465& 0.137939& 0.043948& 0.004171& 0.016915\\
MAD & & &  0.141522& 0.370784& 1.156205& 0.008795& 0.152958\\
\hline\hline
\end{tabular}
\end{table*}

\begin{figure*}[!htbp]
\centering
\begin{subfigure}{0.5\textwidth}
\includegraphics[width=\textwidth]{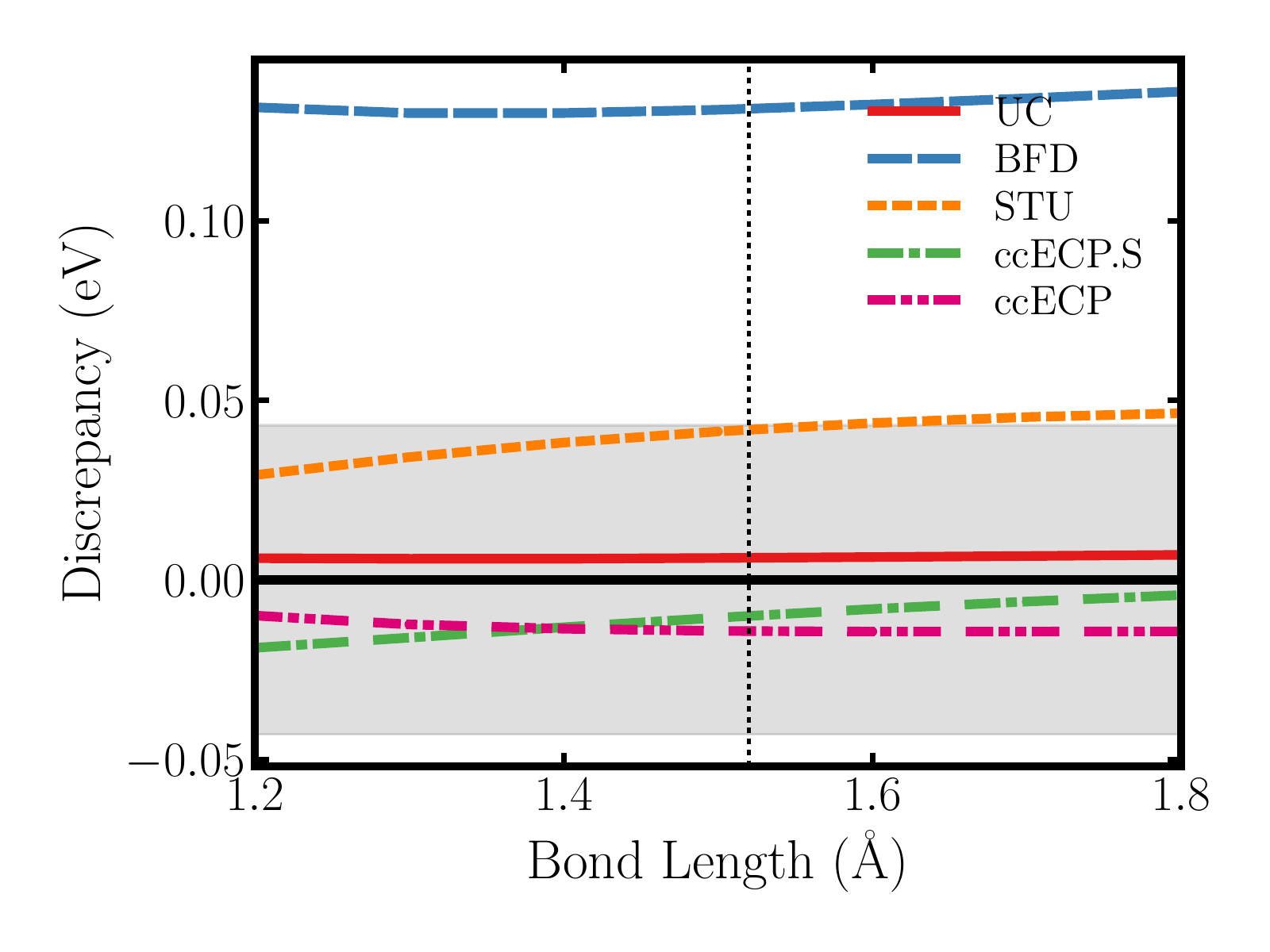}
\caption{CoH binding curve discrepancies.}
\label{fig:CoH}
\end{subfigure}%
\begin{subfigure}{0.5\textwidth}
\includegraphics[width=\textwidth]{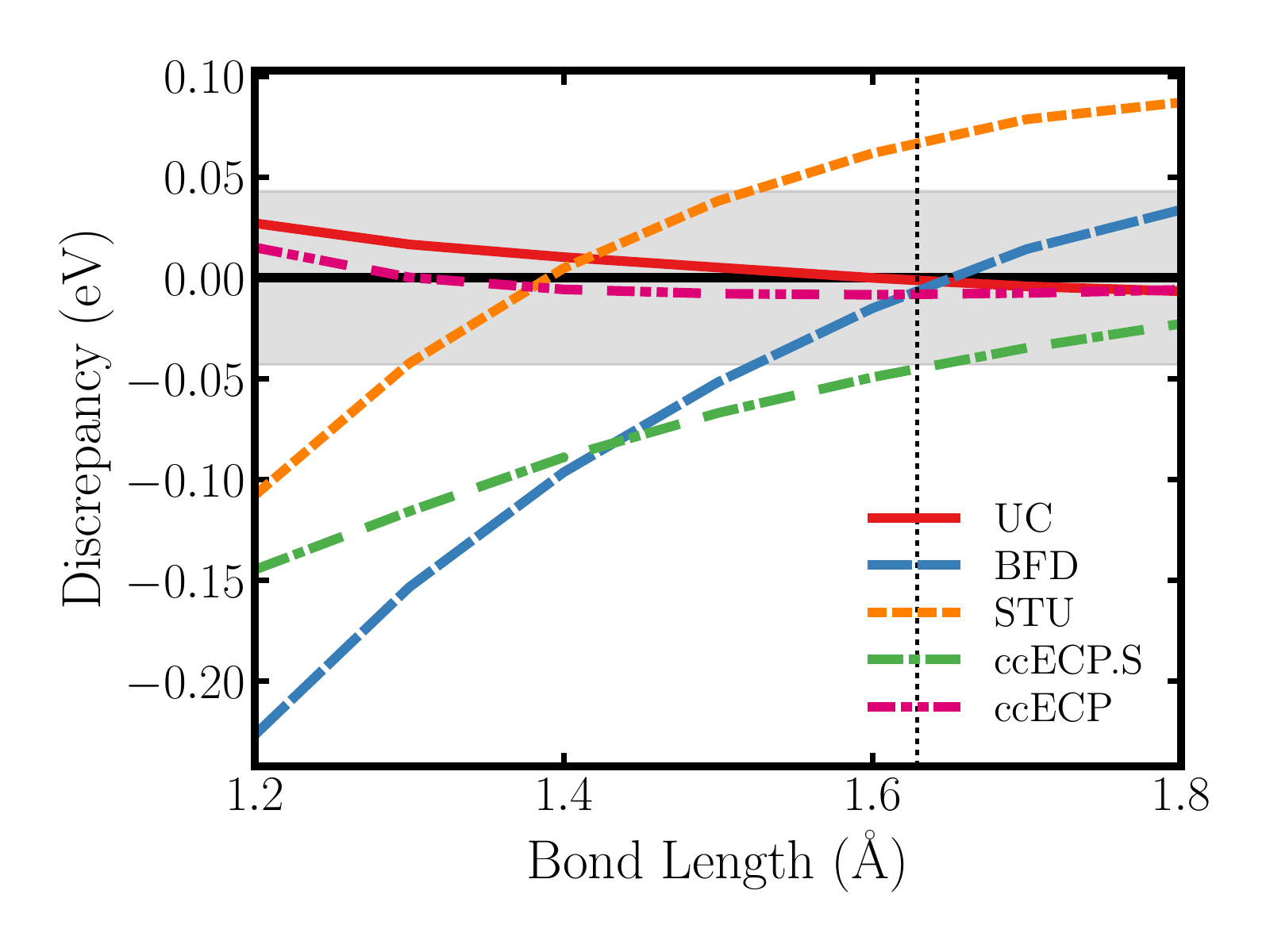}
\caption{CoO binding curve discrepancies}
\label{fig:CoO}
\end{subfigure}
\caption{Binding energy discrepancies for (a) CoH and (b) CoO molecules. The binding curves are relative to the CCSD(T) binding curve. The shaded region indicates a discrepancy of chemical accuracy in either direction.}
\label{fig:Co_mols}
\end{figure*}

The atomic and molecular data for Co are given in Table \ref{tab:Co} and Fig. \ref{fig:Co_mols}, respectively. 
Our spectral construction ccECP.S shows that significant improvements when compared with other core approximations. 
We are able to achieve for both the LMAD and MAD significantly lower discrepancies than UC, BFD, and STU. 
In terms of CoH, the ccECP.S construction looks very satisfactory. 
CoH results in a discrepancy that is well within chemical accuracy for ccECP.S across the entire binding region. 
However, for CoO we see that the spectral construction is less accurate and as in  previous cases it results in significant overbinding. 
Note that other available ECPs have a similar behavior, resulting in overbinding over 0.1~eV near the dissociation threshold.

When we introduce additional constraints into our objective function, see section \ref{Objective}, we are able to obtain a dramatic improvement in the overall transferability. 
With regards to the spectrum, our ccECP is comparable to the UC approximation for the low-lying spectrum (LMAD) and even slightly better for the entire atomic spectrum 
(signaling a complicated landscape of the objective 
function with many minimas). 
For CoH, the binding energy discrepancy is comparable to our ccECP.S, but the overall curve is flatter.
The most profound improvement comes with the  CoO polar bond, where we are able to obtain a nearly flat discrepancy throughout the entire binding region and very marginal 0.01~eV overbinding. 
Note the overall major boost in accuracy when compared to the other ECPs we studied.
The parameters for the ccECP are given in Table \ref{tab:2nd_param}.

\subsection{Nickel}\label{sec:Ni}

\begin{table*}[!htbp]
\centering
\caption{Ni AE gaps and relative errors for various ECPs. All gaps are relative to the [Ar] $3d^8 4s^2$ $^3F$ state. See Tab.\ref{tab:Sc} for further description. All data in eV.}
\label{tab:Ni}
\begin{tabular}{ll|rrrrrrrr}
\hline\hline

Gaps &  &           AE &        UC &       BFD &       STU &   ccECP.S &     ccECP \\
\hline
{[Ar] $3d^94s^2$ } &      $^2D$ &    -1.209029 &  0.005065 &  0.146336 &   0.013536 &   -0.003339 &           0.000304 \\
{[Ar] $3d^94s^1$ } &      $^3D$ &    -0.039999 &  0.010094 &  0.154368 &  -0.007817 &   -0.003776 &          -0.003715 \\
{[Ar] $3d^{10}$ }  &      $^1S$ &     1.665329 &  0.014848 &  0.258005 &   0.022213 &   -0.004163 &          -0.006370 \\
{[Ar] $3d^84s^1$ } &      $^4F$ &     8.678287 &  0.004880 & -0.004271 &  -0.026728 &    0.001234 &          -0.011767 \\
{[Ar] $3d^9$ }     &      $^2D$ &     7.580538 &  0.017010 &  0.152086 &  -0.041737 &    0.002174 &          -0.002632 \\
{[Ar] $3d^8$ }     &      $^3F$ &    25.813278 &  0.013459 & -0.005649 &  -0.039091 &    0.014570 &          -0.007257 \\
{[Ar] $3d^7$ }     &      $^4F$ &    61.040014 & -0.000243 & -0.144218 &   0.100049 &    0.016374 &          -0.042470 \\
{[Ar] $3d^6$ }     &      $^5D$ &   116.199196 & -0.026699 & -0.226084 &   0.435981 &    0.004456 &          -0.105248 \\
{[Ar] $3d^5$ }     &      $^6S$ &   192.239024 & -0.075327 & -0.281178 &   1.016989 &   -0.018535 &          -0.175304 \\
{[Ar] $3d^4$ }     &      $^5D$ &   300.110087 & -0.148578 & -0.080333 &   2.014774 &   -0.012141 &          -0.209976 \\
{[Ar] $3d^3$ }     &      $^4F$ &   433.615540 & -0.256215 &  0.003373 &   3.416095 &    0.010293 &          -0.172026 \\
{[Ar] $3d^2$ }     &      $^3F$ &   596.091441 & -0.391620 & -0.212840 &   5.276018 &    0.023892 &          -0.054807 \\
{[Ar] $3d^1$ }     &      $^2D$ &   789.334381 & -0.559135 & -0.961215 &   7.649629 &    0.010737 &           0.163535 \\
{[Ar] }            &      $^1S$ &  1013.195350 & -0.779566 & -2.452519 &  10.611841 &   -0.013917 &           0.540565 \\
\hline
LMAD & & & 0.010379 & 0.143013 & 0.022406 & 0.002937 & 0.004958\\
MAD & & &  0.164481 & 0.363034 & 2.190893 & 0.009971 & 0.106855\\
\hline\hline
\end{tabular}
\end{table*}

\begin{figure*}[!htbp]
\centering
\begin{subfigure}{0.5\textwidth}
\includegraphics[width=\textwidth]{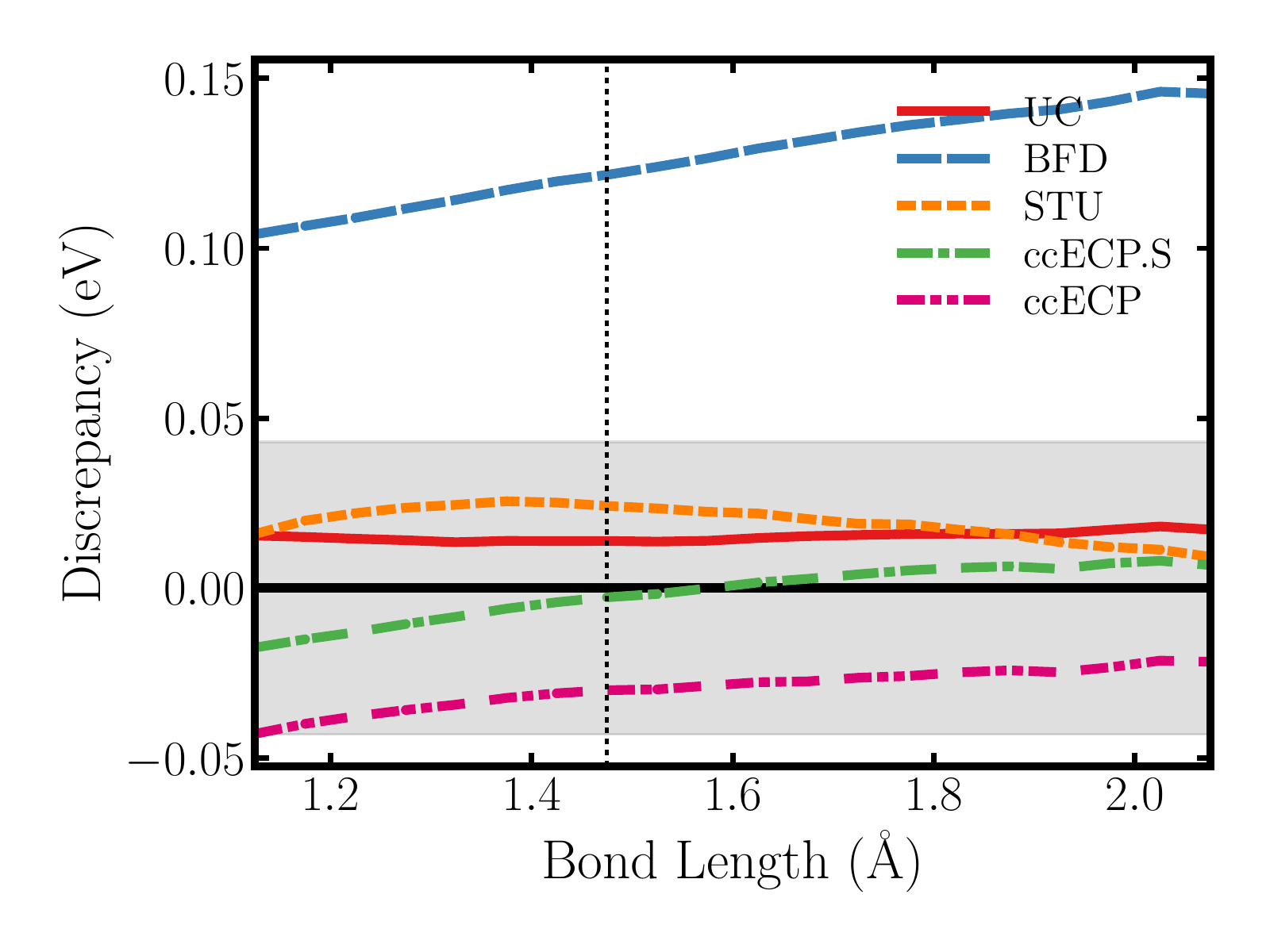}
\caption{NiH binding curve discrepancies.}
\label{fig:NiH}
\end{subfigure}%
\begin{subfigure}{0.5\textwidth}
\includegraphics[width=\textwidth]{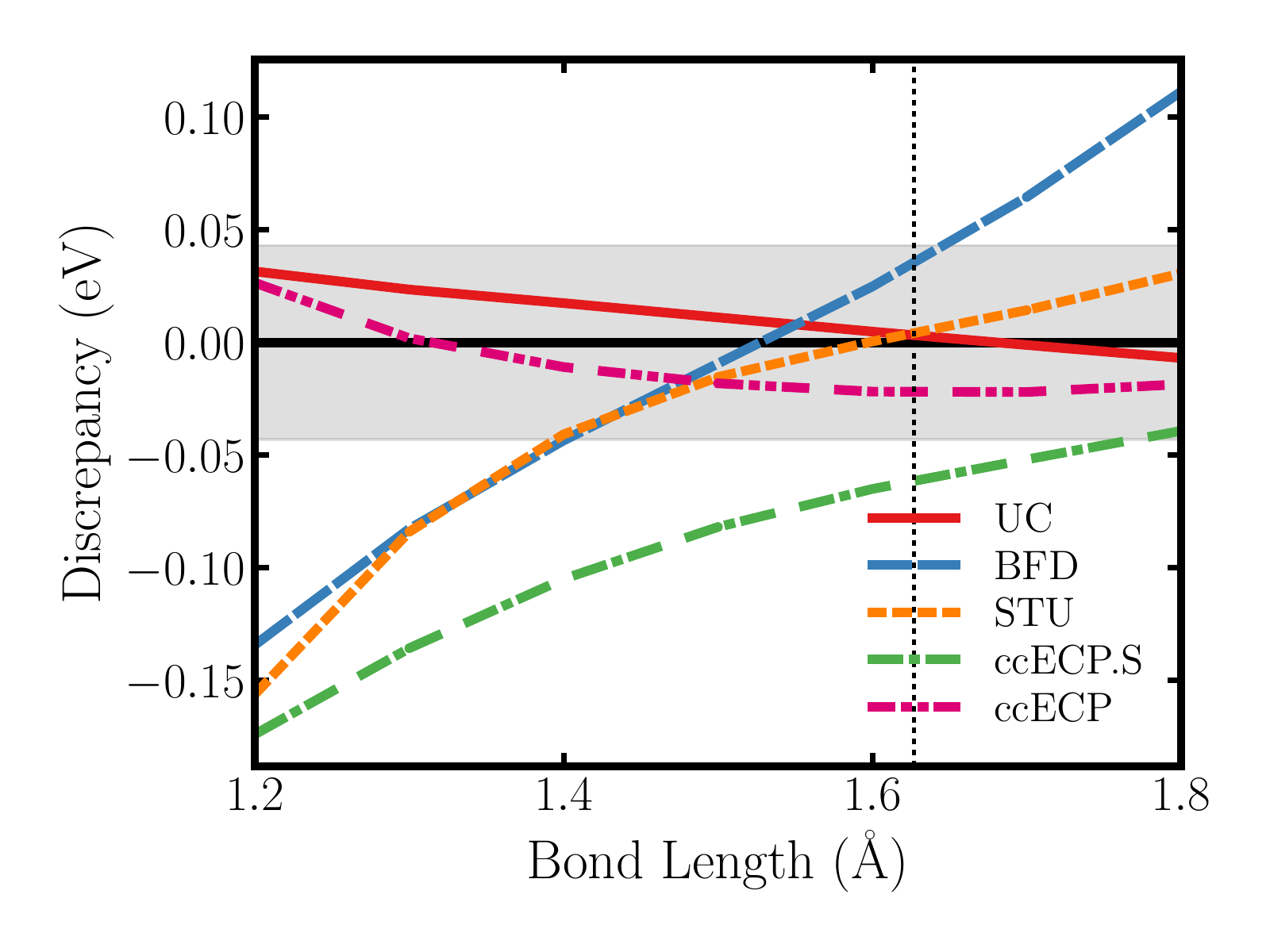}
\caption{NiO binding curve discrepancies}
\label{fig:NiO}
\end{subfigure}
\caption{Binding energy discrepancies for (a) NiH and (b) NiO molecules. The binding curves are relative to the CCSD(T) binding curve. The shaded region indicates a discrepancy of chemical accuracy in either direction.}
\label{fig:Ni_mols}
\end{figure*}

We show the atomic and molecular data for Ni in Table \ref{tab:Ni} and Fig. \ref{fig:Ni_mols}, respectively. 
Optimization for ccECP.S results in an accurate spectrum with a 3~meV LMAD for the low-lying spectrum and 10~meV MAD for the entire spectrum.
These discrepancies are significantly smaller than for the other core approximations of UC, BFD, and STU. 
For NiH, we see significant issues with BFD, whereas all other core-approximations agree with AE to within chemical accuracy. 
We find more sizeable errors for NiO for BFD, STU, and also for our spectral ccECP.S. 

To obtain a more reasonable description of the molecular properties, we reoptimize using Eq. \ref{eqn:final_objective} that produces our ccECP. 
With refinement we compromise on both the LMAD and MAD that mildly increases relative to ccECP.S, however, this increase is caused mainly by the very deeply lying states ($>$ 500~eV) while the low-lying spectrum is still exceptionally well described, much better than for BFD, STU, and comparable to UC. 
While we slightly compromise NiH, we find a significant improvement for NiO, where we reduce the discrepancy to be well within chemical accuracy for all considered bond lengths and maintain a flat discrepancy throughout. 
The final parameters for our ccECP are given in Table \ref{tab:2nd_param}.

\subsection{Copper}\label{sec:Cu}

\begin{table*}[!htbp]
\centering
\caption{Cu AE and relative errors for various core approximations. All gaps are relative to the [Ar] $3d^9 4s^2$ $^2D$ state. See Tab.\ref{tab:Sc} for further description. All data in eV.}
\label{tab:Cu}
\begin{tabular}{ll|rrrrrrrrrrrr}
\hline\hline
Gaps &  &           AE &        UC &       BFD &      STU &     eCEPP &   ccECP.S &      ccECP \\
\hline
 {[Ar] $3d^{10}4s^2$}& $^1S$&      -2.760729 &  0.004101 &  0.053157 &  0.039397 &  -0.000870 &   -0.016530 &         -0.006277 \\
 {[Ar] $3d^{10}4s^1$}& $^2S$&      -1.513939 &  0.007503 &  0.047758 &  0.042300 &   0.004682 &   -0.020428 &         -0.007892 \\
 {[Ar] $3d^9 4s^1$ } & $^3D$&       9.062324 &  0.005137 & -0.006833 &  0.004934 &   0.003888 &   -0.003649 &         -0.002273 \\
 {[Ar] $3d^{10}$ }   & $^1S$&       6.223451 &  0.013241 &  0.014972 &  0.047085 &   0.003103 &   -0.032589 &         -0.008363 \\
 {[Ar] $3d^9$  }     & $^2D$&      26.646257 &  0.012720 & -0.009051 &  0.024636 &   0.003493 &   -0.009803 &          0.010994 \\
 {[Ar] $3d^8$  }     & $^3F$&      63.558190 & -0.000543 & -0.027268 &  0.020881 &   0.003248 &    0.010925 &          0.017731 \\
 {[Ar] $3d^7$  }     & $^4F$&     121.008477 & -0.025039 & -0.028883 &  0.055165 &   0.013036 &    0.018442 &         -0.001066 \\
 {[Ar] $3d^6$  }     & $^5D$&     201.489726 & -0.062319 & -0.024452 &  0.156973 &   0.047778 &    0.010412 &         -0.036229 \\
 {[Ar] $3d^5$   }    & $^6S$&     305.535974 & -0.122740 & -0.058675 &  0.364107 &   0.118660 &   -0.008278 &         -0.054449 \\
 {[Ar] $3d^4$   }    & $^5D$&     444.867210 & -0.207748 &  0.024219 &  0.733521 &   0.331561 &   -0.002246 &         -0.067000 \\
 {[Ar] $3d^3$   }    & $^4F$&     612.257887 & -0.328280 & -0.034713 &  1.317769 &   0.665239 &    0.022817 &          0.031193 \\
 {[Ar] $3d^2$   }    & $^3F$&     811.134546 & -0.476918 & -0.377656 &  2.138293 &   1.124162 &    0.032874 &          0.255770 \\
 {[Ar] $3d^1$   }    & $^2D$&    1043.247944 & -0.657795 & -1.166121 &  3.235735 &   1.709602 &    0.012528 &          0.655110 \\
 {[Ar] }             & $^1S$&    1308.271906 & -0.893022 & -2.520960 &  4.697743 &   2.455655 &   -0.013314 &          1.337623 \\
\hline
LMAD & & & 0.008540& 0.026354& 0.031670& 0.003207& 0.016600& 0.007160 \\
MAD  & & & 0.201222& 0.313908& 0.919896& 0.463213& 0.015345& 0.177998\\
\hline
\end{tabular}
\end{table*}

\begin{figure*}[!htbp]
\centering
\begin{subfigure}{0.5\textwidth}
\includegraphics[width=\textwidth]{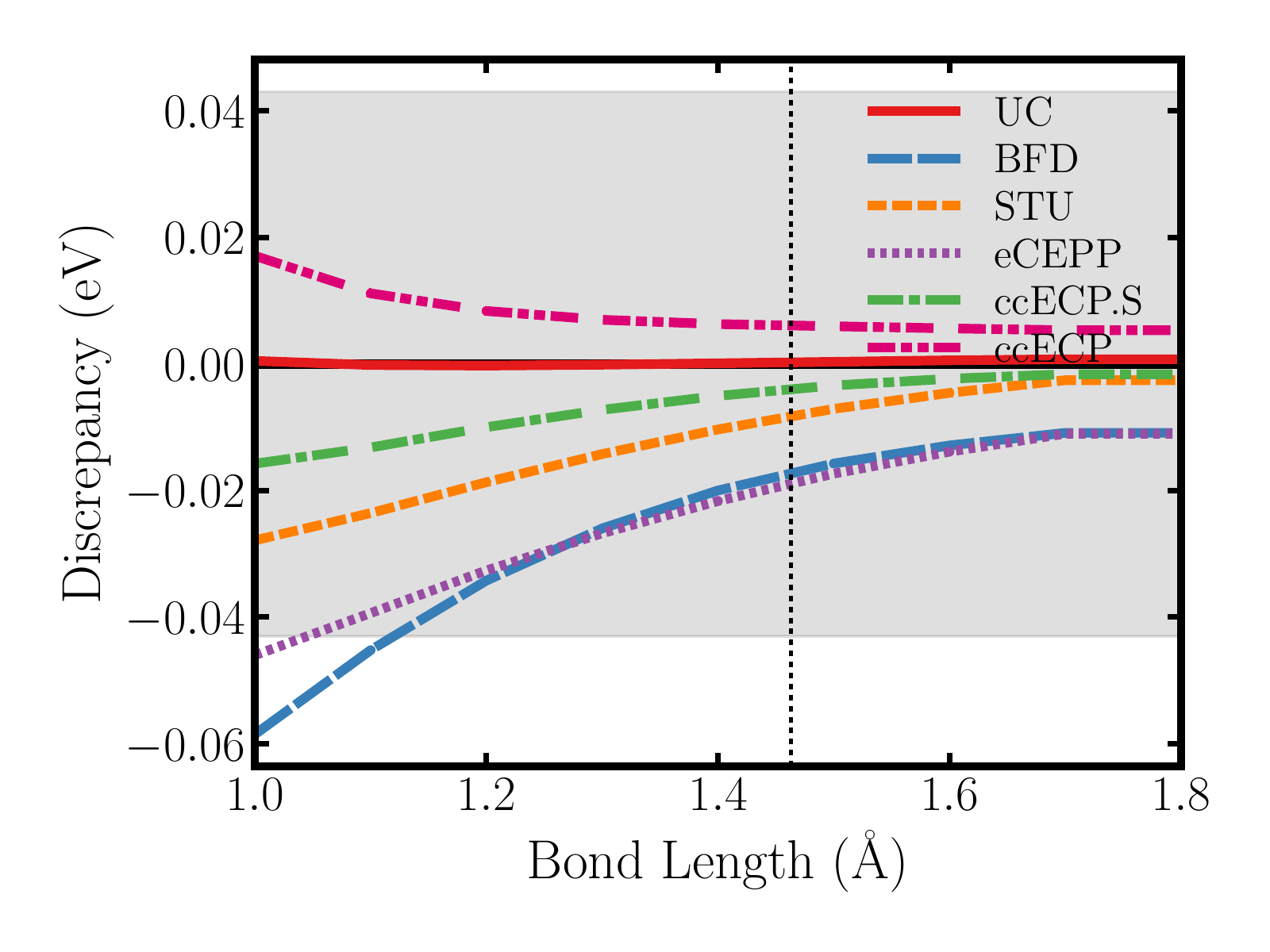}
\caption{CuH binding curve discrepancies.}
\label{fig:CuH}
\end{subfigure}%
\begin{subfigure}{0.5\textwidth}
\includegraphics[width=\textwidth]{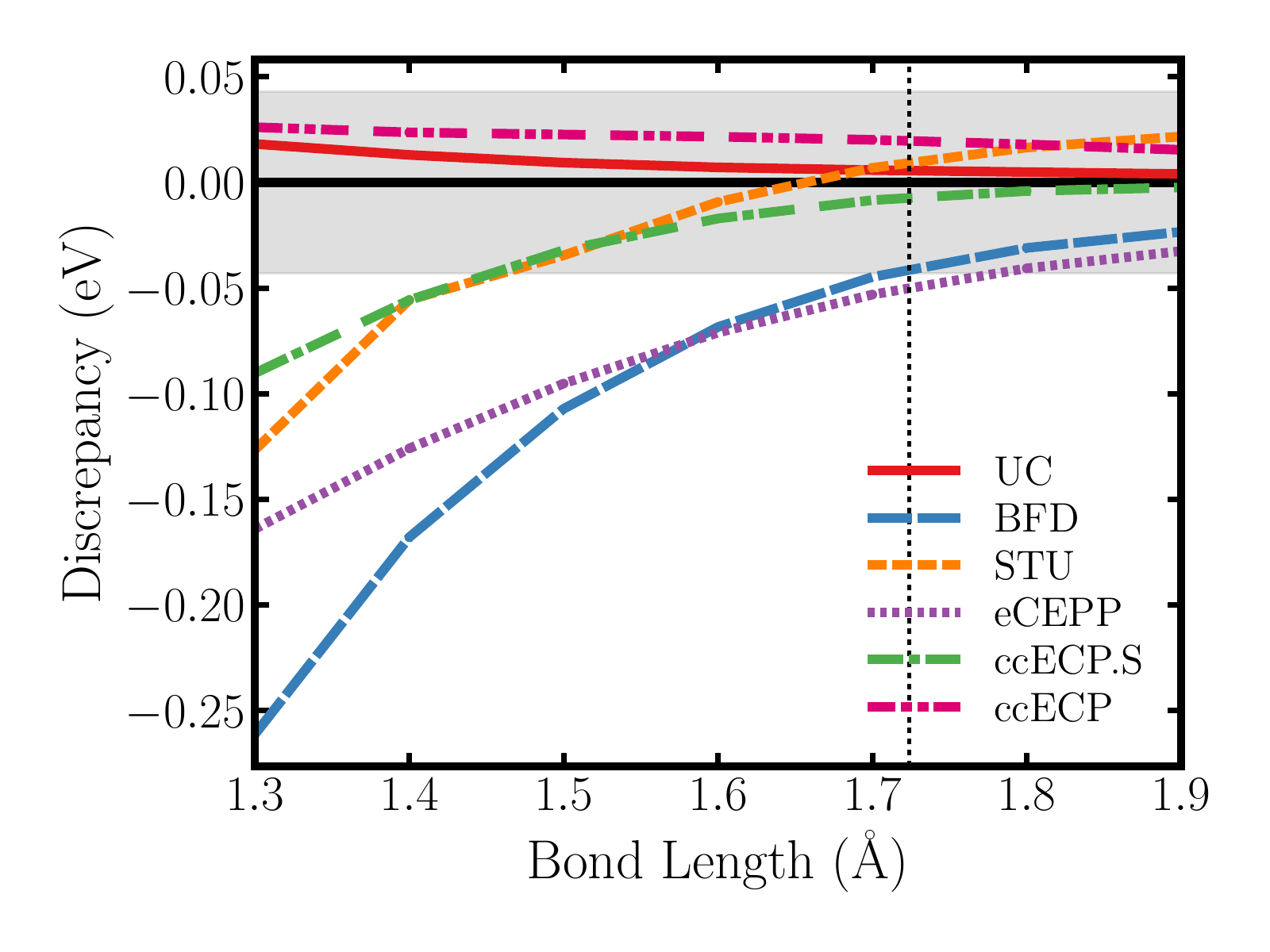}
\caption{CuO binding curve discrepancies}
\label{fig:CuO}
\end{subfigure}
\caption{Binding energy discrepancies for (a) CuH and (b) CuO molecules. The binding curves are relative to the CCSD(T) binding curve. The shaded region indicates a discrepancy of chemical accuracy in either direction.}
\label{fig:Cu_mols}
\end{figure*}

We show the atomic properties for Cu in Table \ref{tab:Cu} and the molecular properties for CuH and CuO in Fig. \ref{fig:Cu_mols}.
Note that this table shows gaps with respect to [Ar]$3d^9 4s^2$ $^2D$ although the ground state is [Ar]$3d^{10} 4s^1$ $^2S$, to be consistent with the other elements. 
Due to the closed d-shell LMAD includes 4 excitations, whereas previous tables for Sc-Ni included 5.  
Focusing first on the spectral ccECP.S, we are able to obtain an almost uniform agreement for the entire spectrum, with a LMAD of 0.018~eV and a MAD of 0.015~eV. 
While the LMAD is slightly higher than UC and eCEPP, the overall MAD is significantly improved since the high ionizations are in much better agreement to AE.
As has been seen for the previous elements, a uniform accuracy on the atomic spectrum can decrease the overall transferability. 
While CuH is quite well described by the ccECP.S, we again find significant overbinding near dissociation for our ccECP.S.
In fact, {\em all ECPs} including STU, BFD, and eCEPP significantly overbind for the CuO molecule, from roughly 0.1~eV for our ccECP.S to over 0.25~eV for BFD. 

We substantially improve the overall transferability with our refined ccECP. 
For both CuH and CuO, we have near perfect agreement with AE along the entire binding curve.
For CuO, we are able to obtain the accuracy 
that stands out when compared to previous ECPs. 
Note that behavior of biases for this element suggests that keeping the accuracy for energies  $>500$ eV one would need more variational freedom and perhaps a more sophisticated construction. This is due to different energy and also length scales for semicore vs valence spaces. 

\subsection{Zinc}\label{sec:Zn}

\begin{table*}[!htbp]
\caption{Zn AE gaps and relative errors for various ECPs. All gaps are relative to the [Ar] $3d^{10} 4s^2$ $^1S$ state. See Tab.\ref{tab:Sc} for further description. All data in eV.}
\label{tab:Zn}
\begin{tabular}{ll|rrrrrrrrrr}
\hline
Gaps & &           AE &        UC &       BFD &       STU &   ccECP.S &     ccECP \\
\hline
{[Ar] $3d^{10}4s^1$ } &      $^2S$ &     9.411709 &  0.004952 &  0.018765 &  0.004463 &  0.007538 &  0.000215 \\
{[Ar] $3d^94s^2$ }    &      $^2D$ &    17.323005 & -0.009362 & -0.205955 & -0.022585 &  0.009686 &  0.022019 \\
{[Ar] $3d^{10}$ }     &      $^1S$ &    27.392261 &  0.011411 &  0.036333 &  0.006989 &  0.002052 & -0.009078 \\
{[Ar] $3d^9$ }        &      $^2D$ &    67.262367 &  0.003722 & -0.050548 &  0.003788 &  0.010425 &  0.013184 \\
{[Ar] $3d^8$ }        &      $^3F$ &   126.975193 & -0.018903 & -0.090639 &  0.051139 &  0.013474 &  0.028731 \\
{[Ar] $3d^7$ }        &      $^4F$ &   210.344966 & -0.052708 & -0.055643 &  0.173830 &  0.006980 &  0.031967 \\
{[Ar] $3d^6$ }        &      $^5D$ &   319.614848 & -0.099359 &  0.033786 &  0.406930 & -0.008892 &  0.026004 \\
{[Ar] $3d^5$ }        &      $^6S$ &   455.004148 & -0.170049 &  0.098292 &  0.795761 & -0.026917 &  0.025026 \\
{[Ar] $3d^4$ }        &      $^5D$ &   629.026722 & -0.264676 &  0.351989 &  1.415417 & -0.015099 &  0.072364 \\
{[Ar] $3d^3$ }        &      $^4F$ &   833.485035 & -0.396401 &  0.369407 &  2.316968 &  0.018092 &  0.171157 \\
{[Ar] $3d^2$ }        &      $^3F$ & 1,071.906266 & -0.556213 & -0.076598 &  3.529685 &  0.034067 &  0.296417 \\
{[Ar] $3d^1$ }        &      $^2D$ & 1,346.006047 & -0.748315 & -2.414393 &  5.101447 &  0.007904 &  0.438073 \\
{[Ar]}                &      $^1S$ & 1,655.288042 & -0.996097 & -3.118490 &  7.132298 & -0.032440 &  0.638571 \\
\hline
LMAD & & & 0.007157 & 0.112360 & 0.013524 & 0.008612 & 0.011117 \\
MAD & & & 0.256321 & 0.532372 & 1.612408 & 0.014890 & 0.136370 \\
\hline\hline
\end{tabular}
\end{table*}

\begin{figure*}[!htbp]
\centering
\begin{subfigure}{0.5\textwidth}
\includegraphics[width=\textwidth]{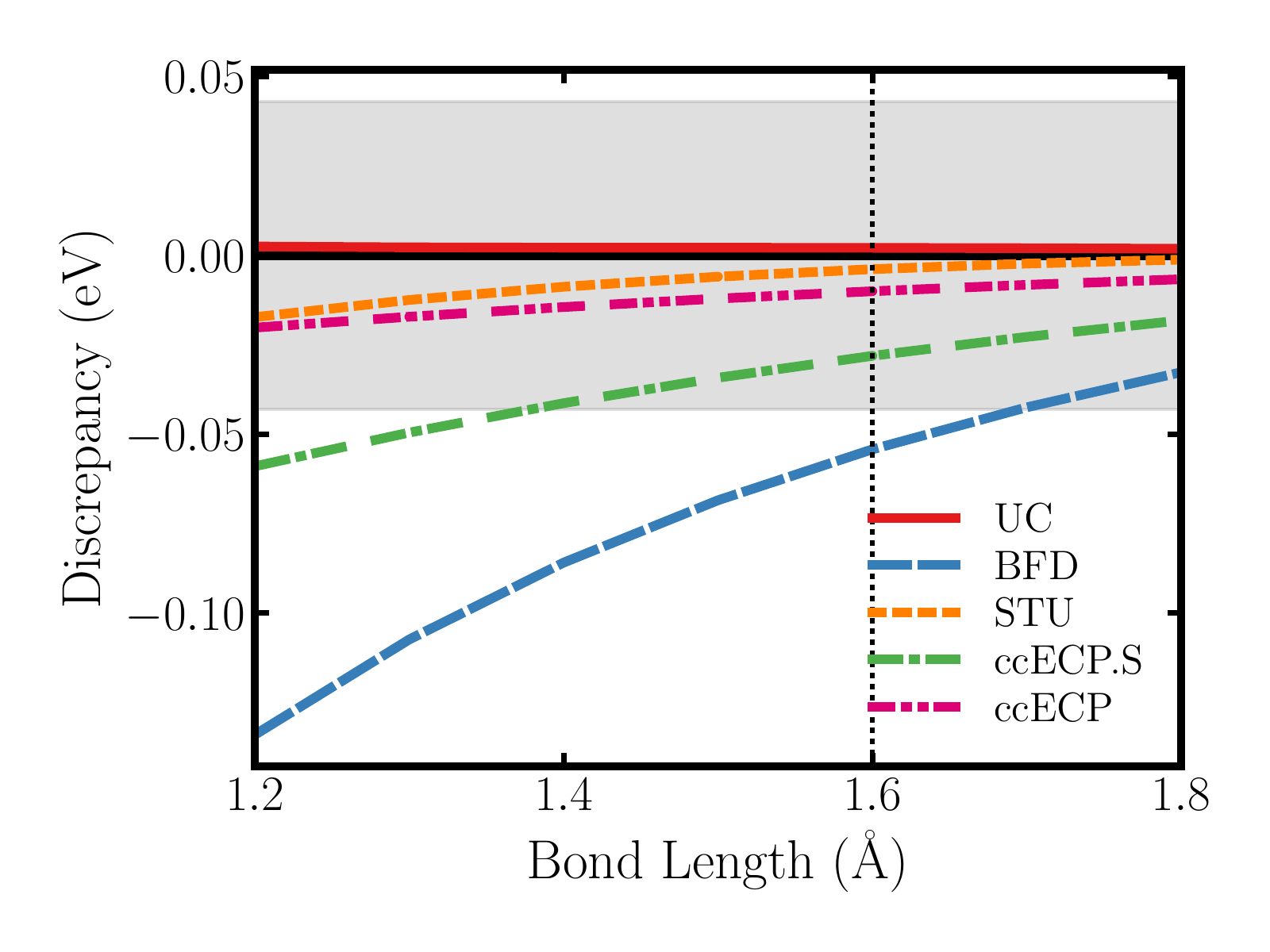}
\caption{ZnH binding curve discrepancies.}
\label{fig:ZnH}
\end{subfigure}%
\begin{subfigure}{0.5\textwidth}
\includegraphics[width=\textwidth]{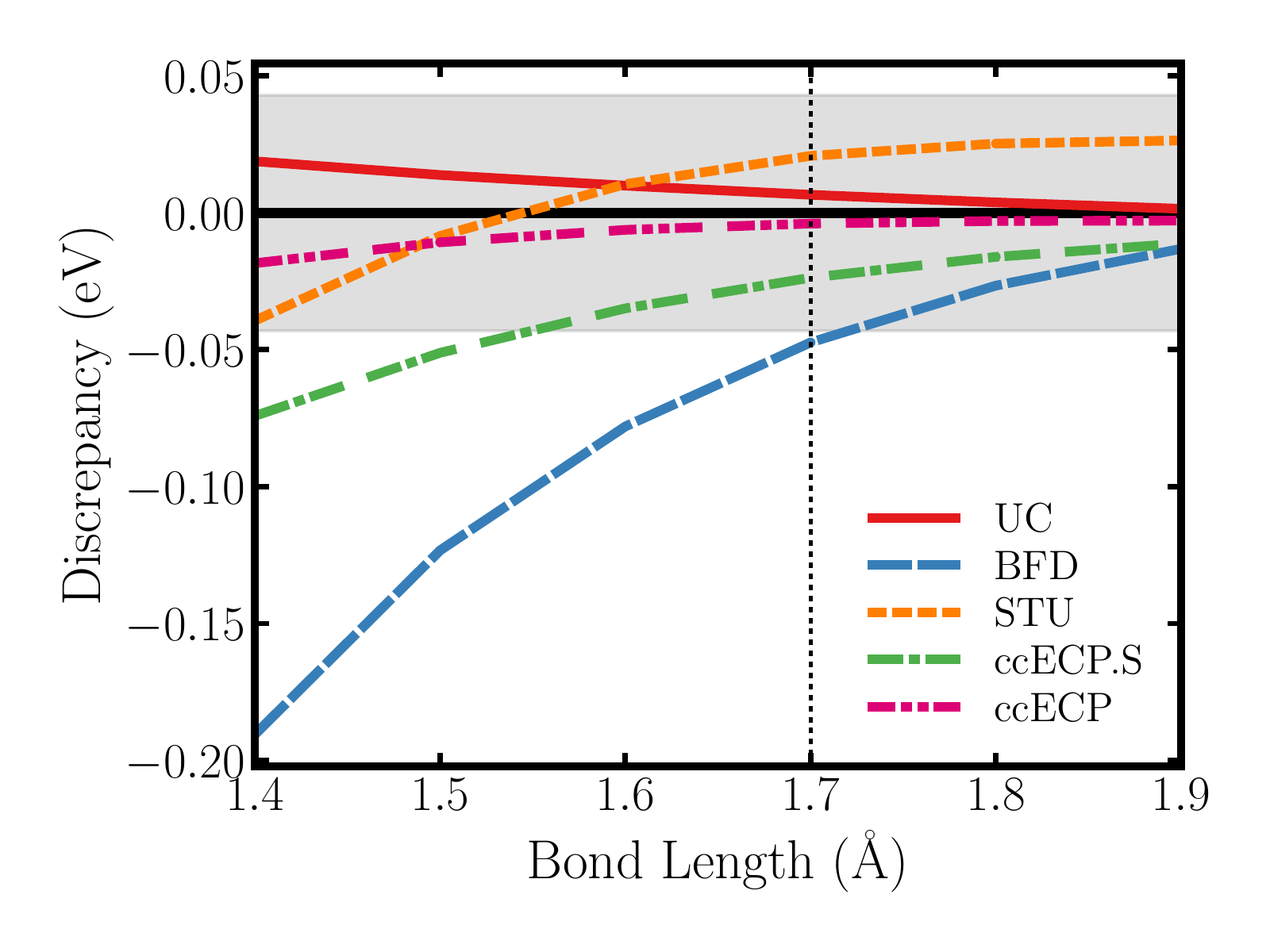}
\caption{ZnO binding curve discrepancies}
\label{fig:ZnO}
\end{subfigure}
\caption{Binding energy discrepancies for (a) ZnH and (b) ZnO molecules. The binding curves are relative to the CCSD(T) binding curve. The shaded region indicates a discrepancy of chemical accuracy in either direction.}
\label{fig:Zn_mols}
\end{figure*}

The atomic and molecular data for Zn is given in Table \ref{tab:Zn} and Fig. \ref{fig:Zn_mols}, respectively. 
For our spectral ccECP.S, we are able to achieve a uniform accuracy across the entire atomic spectrum, resulting in a MAD of 0.015~eV. 
This outperforms BFD, STU, and UC significantly, resulting in an ECP that
on spectrum is very close to the original AE Hamiltonian. 
Additionally, the LMAD for Zn includes only the first two states, due to the fact that the anion and most of the s$\leftrightarrow $d transitions do not exist.
For the transferability tests, we find similar behavior for both ZnH and ZnO, where the ccECP.S is within chemical accuracy near equilibrium but begins to overbind as the bond is compressed. 

With the refinements from our ccECP, we are able to strike a reasonable balance between the atomic properties as well as the molecular transferability. 
In both the hydride and oxide, the ccECP has a relatively flat discrepancy resulting in excellent vibrational frequencies as well as energy differences well within chemical accuracy for the entire curve.
In terms of the atomic spectrum, the lowest states are only 0.01~eV error and we achieve a MAD on the entire spectrum that is better than all other core approximations. 
The parameters for our final ccECP are given in Table \ref{tab:2nd_param}.

\begin{table}[!htbp]
    \centering
    \caption{Parameter values for hydrogen. The highest $l$ value corresponds to the local channel. Each term takes the form $V_{\ell k}(r) = \beta_{\ell k}r^{n_{\ell k} -2}\exp\left[ -\alpha_{\ell k} r^2\right]$}
    \begin{tabular}{ccccrr}
	\hline\hline
         Atom &   $Z_{\mathrm{eff}}$ &  $\ell$ & $n_{\ell k}$ &  \multicolumn{1}{c}{$\alpha_{\ell k}$} & \multicolumn{1}{c}{$\beta_{\ell k}$} \\
     \hline
          &             &      &     &              &              \\   
       H  &         1   &  0   &   2 &  1.00000000  &   0.00000000 \\
          &             &  1   &   1 & 21.24359508  &   1.00000000 \\
          &             &  1   &   3 & 21.24359508  &  21.24359508 \\
          &             &  1   &   2 & 21.77696655  & -10.85192405 \\
          &             &      &     &              &              \\   
	\hline\hline
    \end{tabular}
    \label{tab:pp_H}
\end{table}

\subsection{Average molecular discrepancies }\label{sec:Mol}

In Tab. \ref{tab:mol_mean} we collect the results of molecular calculations for all elements and evaluate mean 
absolute deviations for the equilibrium molecular parameters. Note that even our ccECP.S set achieves very respectable accuracy that is comparable or better than the available tables. Clearly, the adjusted set ccECP is the most balanced overall, to the best of our knowledge shows overall the best consistency for all the calculated parameters. 

\begin{table*}[!htbp]
\centering
\caption{Mean absolute deviations of binding parameters for various core approximations with respect to AE data for transition metal hydride and oxide molecules. All parameters were obtained using Morse potential fit. The parameters shown are dissociation energy $D_e$, equilibrium bond length $r_e$, vibrational frequency $\omega_e$ and binding energy discrepancy at dissociation bond length $D_{diss}$.}
\label{tab:mol_mean}
\begin{tabular}{l|rrrrrr}
\hline\hline
{} &               UC &              BFD &              STU &            eCEPP &          ccECP.S &            ccECP \\
\hline
$D_e$(eV)             &  0.0065(40) &  0.0592(41) &  0.0382(41) &  0.0163(45) &  0.0238(40) &  0.0117(40) \\
$r_e$(\AA)            &  0.0012(13) &  0.0064(13) &  0.0026(13) &  0.0019(15) &  0.0027(13) &  0.0010(13) \\
$\omega_e$(cm$^{-1}$) &    2.3(5.8) &   10.4(6.0) &    4.6(5.9) &    3.9(6.9) &    6.4(5.8) &    3.0(5.8) \\
$D_{diss}$(eV)        &   0.021(41) &   0.146(41) &   0.035(41) &   0.032(46) &   0.054(40) &   0.018(41) \\
\hline\hline
\end{tabular}
\end{table*}


\onecolumngrid

\begin{table*}[htbp!]
    \centering
    \small
    \begin{minipage}{0.5\linewidth}
	\caption{Parameter values for early transition metal ECPs. For all ECPs, the highest $l$ value corresponds to the local channel. Each term takes the form $V_{\ell k}(r) = \beta_{\ell k}r^{n_{\ell k} -2}\exp\left[ -\alpha_{\ell k} r^2\right]$}%
    \begin{tabular}{ccccrr}
	 \hline
	 \hline
     Atom &   $Z_{\mathrm{eff}}$ &  $\ell$ & $n_{\ell k}$ &  \multicolumn{1}{c}{$\alpha_{\ell k}$} & \multicolumn{1}{c}{$\beta_{\ell k}$} \\
     \hline
          &             &      &     &              &               \\   
       Sc &          11 &   0  &  2  &  11.49466541 &  153.96530175 \\  
          &             &   0  &  2  &   5.01031394 &   14.93675657 \\
          &             &   1  &  2  &  11.45126730 &   97.21725690 \\
          &             &   1  &  2  &   4.76798446 &   10.81704018 \\
          &             &   2  &  1  &  16.02394388 &   11.00000000 \\
          &             &   2  &  3  &  14.08647403 &  176.26338271 \\
          &             &   2  &  2  &  11.93985121 &  -83.68149599 \\
          &             &   2  &  2  &   3.69440111 &   0.432827647 \\
          &             &      &     &              &               \\    
       Ti &          12 &   0  &  2  &  12.70580614 &  173.94657236 \\
          &             &   0  &  2  &   6.11178552 &   18.83768334 \\
          &             &   1  &  2  &  12.64091930 &  111.45672882 \\
          &             &   1  &  2  &   5.35437416 &   11.17702683 \\
          &             &   2  &  1  &  18.41366202 &   12.00000000 \\
          &             &   2  &  3  &  15.92292414 &  220.96394426 \\
          &             &   2  &  2  &  13.65000623 &  -94.29025825 \\
          &             &   2  &  2  &   5.09555211 &    0.09791142 \\
          &             &      &     &              &               \\    
       V  &          13 &   0  &  2  &  15.12502151 &  195.56713891 \\
          &             &   0  &  2  &   6.29898914 &   22.88642835 \\
          &             &   1  &  2  &  15.93855113 &  126.42119501 \\
          &             &   1  &  2  &   5.74006267 &   16.03597128 \\
          &             &   2  &  1  &  20.32168914 &   13.00000000 \\
          &             &   2  &  3  &  19.59698040 &  264.18195885 \\
          &             &   2  &  2  &  17.33147348 & -115.29293208 \\
          &             &   2  &  2  &   5.12320658 &   -0.66288726 \\
          &             &      &     &              &               \\    
       Cr &          14 &   0  &  2  &  16.90078761 &  219.48146210 \\
          &             &   0  &  2  &   7.33662151 &   28.07933177 \\
          &             &   1  &  2  &  17.31974517 &  139.98396872 \\
          &             &   1  &  2  &   6.92409758 &   19.54835786 \\
          &             &   2  &  1  &  18.28091074 &   14.00000000 \\
          &             &   2  &  3  &  17.09800655 &  255.93275041 \\
          &             &   2  &  2  &  16.72267276 & -132.01826317 \\
          &             &   2  &  2  &   5.02865106 &   -0.77388761 \\
          &             &      &     &              &               \\    
       Mn &          15 &   0  &  2  &  18.92044966 &  244.66870493 \\   
          &             &   0  &  2  &   8.32764757 &   33.54162717 \\
          &             &   1  &  2  &  20.17347020 &  162.35033686 \\
          &             &   1  &  2  &   7.80047874 &   24.17956695 \\
          &             &   2  &  1  &  21.91937433 &   15.00000000 \\
          &             &   2  &  3  &  21.35527128 &  328.79061500 \\
          &             &   2  &  2  &  21.27162654 & -162.05172805 \\
          &             &   2  &  2  &   7.93913962 &   -1.82694273 \\
	  &             &      &     &              &               \\    
      
	 \hline
	 \hline
    \end{tabular}\label{tab:1st_param}
    \end{minipage}%
    \begin{minipage}{0.5\linewidth}
    \caption{Parameter values for late transtion metal ECPs. For all ECPs, the highest $l$ value corresponds to the local channel. Each term takes the form $V_{\ell k}(r) = \beta_{\ell k}r^{n_{\ell k} -2}\exp\left[ -\alpha_{\ell k} r^2\right]$}%
    \begin{tabular}{ccccrr}
	 \hline
	 \hline
     Atom &   $Z_{\mathrm{eff}}$ &  $\ell$ & $n_{\ell k}$ &  \multicolumn{1}{c}{$\alpha_{\ell k}$} & \multicolumn{1}{c}{$\beta_{\ell k}$} \\
     \hline
          &             &      &     &              &               \\   
       Fe &          16 &   0  &  2  &  22.21062697 &  277.50032547 \\
          &             &   0  &  2  &   9.51515801 &   46.20495585 \\
          &             &   1  &  2  &  24.57000872 &  194.99875057 \\
          &             &   1  &  2  &   8.86648777 &   31.67945133 \\
          &             &   2  &  1  &  23.22091714 &   16.00000000 \\
          &             &   2  &  3  &  23.54714680 &  371.53467418 \\
          &             &   2  &  2  &  23.47256345 & -181.22603445 \\
          &             &   2  &  2  &   9.85238815 &   -2.37305236 \\
          &             &      &     &              &               \\    
       Co &          17 &   0  &  2  &  23.41427031 &  271.77708487 \\
          &             &   0  &  2  &  10.76931694 &   54.26461122 \\
          &             &   1  &  2  &  25.47446317 &  201.53430745 \\
          &             &   1  &  2  &  10.68404901 &   38.99231927 \\
          &             &   2  &  1  &  25.00124116 &   17.00000000 \\
          &             &   2  &  3  &  22.83490097 &  425.02109972 \\
          &             &   2  &  2  &  23.47468156 & -195.48211283 \\
          &             &   2  &  2  &  10.33794825 &   -2.81572866 \\
          &             &      &     &              &               \\    
       Ni &          18 &   0  &  2  &  13.55641859 &   92.06283029 \\  
          &             &   0  &  2  &  27.83578932 &  332.58121624 \\
          &             &   1  &  2  &   6.49273676 &    7.52030073 \\
          &             &   1  &  2  &  23.68135451 &  266.09985511 \\
          &             &   2  &  1  &  37.92402538 &   18.00000000 \\
          &             &   2  &  3  &  23.91559743 &  682.63245691 \\
          &             &   2  &  2  &  19.97959496 & -173.86371303 \\
          &             &   2  &  2  &   3.59972497 &  0.3662705552 \\
          &             &      &     &              &               \\    
       Cu &          19 &   0  &  2  &  29.35562243 &  370.71371825 \\
          &             &   0  &  2  &  12.77235920 &   66.27560813 \\
          &             &   1  &  2  &  33.51694544 &  271.66281028 \\
          &             &   1  &  2  &  12.52471485 &   49.76265057 \\
          &             &   2  &  1  &  31.53811263 &   19.00000000 \\
          &             &   2  &  3  &  31.06925531 &  599.22413998 \\
          &             &   2  &  2  &  30.59035868 & -244.68915484 \\
          &             &   2  &  2  &  14.05141064 &   -1.29349526 \\
          &             &      &     &              &               \\    
       Zn &          20 &   0  &  2  &  35.02141357 &  431.70804303 \\
          &             &   0  &  2  &  14.63498692 &   95.87640437 \\
          &             &   1  &  2  &  42.22979235 &  313.57770564 \\
          &             &   1  &  2  &  14.57429304 &   74.01270049 \\
          &             &   2  &  1  &  35.80797616 &   20.00000000 \\
          &             &   2  &  3  &  34.53646084 &  716.15952324 \\
          &             &   2  &  2  &  28.62830178 & -204.68393324 \\
          &             &   2  &  2  &   7.96239683 &    0.76026614 \\
          &             &      &     &              &               \\    
          
	 \hline
	 \hline
    \end{tabular}\label{tab:2nd_param}
    \end{minipage}
    \label{tab:ecp_params}%
\end{table*}

\twocolumngrid

\clearpage

\section{Conclusions}\label{Conclusions}
In this paper, we present the correlation consistent ECPs
for 3d elements. 
First, we constructed the ECPs 
using spectral optimizations for all the elements. This
optimization was iterative and included states that 
ionize the given atom down to the [Ne]$3s^2$ ion, as well as 
s$\leftrightarrow$d transfer energies for low-lying atomic and ionic states, including bounded anions. Quite unexpectedly, for most atoms, we were able to optimize the states within 0.01 eV or even smaller 
discrepancies for the full span of ionization energies,
sometimes with accuracies better than the known 
spectroscopic data and also better than the inherent systematic biases in our methodology. As discussed above any discrepancies smaller than $\approx$ 0.05 eV
are in fact comparable to estimated systematic errors in our calculations.
These spectral-only results simply pointed out that such 
fits are indeed feasible using rather small number of free ECP parameters. 

We further probed for accuracy and transferability 
of the constructed operators on transition metal-hydride and oxide molecules. The hydrides 
appear mostly less problematic and
within the error bar window of $\pm$0.05 eV
for all bond lengths with a few exceptions as analyzed above.
The oxide molecules and their polar bonds revealed more complicated picture and most 
ECPs required refinement that has led to high fidelity and chemical accuracy along the binding curves. That set is labeled as the recommended ccECP. The spectral-only ccECP.S versions could still be  useful for atomic calculations  where the accuracy of highly ionized states would be of crucial interest. We note that in cases of Sc and 
Ti ccECP=ccECP.S since the spectral optimization provided desired accuracy without further refinement.  
It is quite remarkable that such accurate 
fits can be constructed using just 
the spectral information as an input.

For high spin elements, the spectral-only ECPs, while still respectable, were less accurate then desired. For geometries around the equilibrium and the stretched bond regions were described 
 very well, however, for shorter bond length regions we observed in some cases overbinding of the order $\approx$ 0.1 eV.  Although 
rather small in a relative sense we have opted for further improvement that took into account 
the decreasing importance of correlations of deeply ionized states vs. the key desired accuracy of the low-lying excitations most relevant for valence properties.
Adding a shift towards the HF
eigenvalues into the objective function enabled us to
mildly retune the 3s and 3p channels and 
that proved to be sufficient to get the binding curves within the 0.05 eV accuracy threshold. This caused some increase in spectral errors, however, only for very 
highly ionized states that are of the order of $> 500$ eV  with relative errors 
being still very small ($\approx$ 0.08\% or smaller).

For the late transition elements namely, Co, Ni, Cu, and Zn we have observed similar behavior with an overbinding tendency for 
oxide molecules at short bond lengths. Similar refinement as for the high-spin elements 
have enabled to alleviate these deviations
so that all the binding curves were within 
the chemical accuracy threshold. For these elements we also note additional 
complications with the largest basis sets for Co, 
Cu and Zn where we have encountered 
the feasibility limits of the used codes.
Therefore we have restricted ourselves to the 
best results we could obtain with reasonable computational  resources and we consider the achieved systematic 
comparisons as adequate for our present purposes.
 
Together with accuracy tests, we have also tried to estimate
the exact atomic total energies for both the ground 
and excited states. The extrapolations from
 extensive basis sets aug-cc-pCV$n$Z with $n$=2,3,4,5 provide rather a consistent picture of the corresponding correlation energies, however, at present  we do not  claim better systematic accuracy than
$\approx$ 10 mHa. In order to 
provide these energies with better 
uncertainties, we plan further investigation in a subsequent study that will be devoted to decrease the error margins to much lower values. 


An observation on the legacy ECP constructions is that
our results confirm that the Stuttgart-Koeln-Bonn 
ECPs \cite{STU:1987, DolgCao} established and derived by Stoll, Dolg and coworkers over the past three decades show systematic
consistency and respectable accuracy for the 3d series.  The discrepancies are mostly  within the 0.1
- 0.2 eV
margins for low-lying atomic excitations and also for most of the molecular binding curves. Since the STU table was constructed with Dirac-Hartree-Fock spectral fits, this confirms that the precise self-consistent energy differences data 
is the dominant factor in achieving consistent 
behavior. However,  correlation 
effects become important at finer
resolutions that we have targeted in this work.

Clearly, our present constructions 
raise the bar of accuracy
higher and we believe that they provide 
a significant step forward for correlated calculations. Further data including the basis sets, including both cc-pV$n$Z and aug-cc-pV$n$Z for $n \in \{D,T,Q,5\}$, can be found 
in web library \url{http://pseudopotentiallibrary.org} as well as in the Supplementary
Information.
Data for this work is also provided at the Materials Data Facility. 


The presented results show that there is a room for sizeable improvements of both the construction methods and practical 
versions of ECPs. In agreement with previous papers on first and second rows, we were aiming 
at offering a "minimal model" that
is still more accurate than existing tables. For example,
nonlocal s and p channels are described by two gaussians only
(similarly to the STU table). We have demonstrated that such a combination of small variational space and accuracy is indeed feasible and the resulting constructions are ready for general use.  
There are several directions 
where this work could expand further. 
Heavier atoms such as Co, Cu and 
Zn could benefit from more variational parameters that would 
address differences between deeply lying 3s, 3p states vs large valence subshell. 
In addition, more testing and validation is needed in a variety of
chemical systems such as larger molecules, solids and 2D materials. Indeed, we expect 
that such data would provide new insights and possibly point out the directions for further refinements and updates in future.

\bigskip

{\bf Acknowledgments.}

We are grateful to P. Kent for reading the manuscript and for comments. This work has been supported by 
 the U.S. Department of Energy, Office of Science, Basic Energy Sciences, Materials Sciences and Engineering Division, as part of the Computational Materials Sciences Program and Center for Predictive Simulation of Functional Materials.

The calculations were performed mostly at Sandia National Laboratories and  at TACC under XSEDE.

Sandia National Laboratories is a multimission laboratory managed and operated by National Technology and Engineering Solutions of Sandia LLC, a wholly owned subsidiary of Honeywell International Inc. for the U.S. Department of Energy's National Nuclear Security Administration under contract DE-NA0003525.

\bibliography{main.bib}

\end{document}